\documentclass[final,3p,times]{elsarticle}

\usepackage{graphicx}
\usepackage{amssymb}
\usepackage{amsmath}
\usepackage{color}
\usepackage{dcolumn}
\usepackage{bm}
\usepackage{xspace}
\usepackage{subfigure}

\newcommand{\ii}{\mathrm{i}}

\newcommand{\TCC}{\ensuremath{\mathrm{TlCuCl_3}}\xspace}
\newcommand{\TCX}{\ensuremath{\mathrm{Tl_xK_{(1-x)}CuCl_3}}\xspace}
\newcommand{\PHCX}{\ensuremath{\mathrm{(C_4H_{12}N_2)Cu_2(Cl_{(1-x)}Br_{x})_6}}\xspace}
\newcommand{\PHCC}{\ensuremath{\mathrm{(C_4H_{12}N_2)Cu_2Cl_6}}\xspace}
\newcommand{\IPA}{\ensuremath{\mathrm{(CH_3)_2CHNH_3CuCl_3}}\xspace}
\newcommand{\IPAX}{\ensuremath{\mathrm{(CH_3)_2CHNH_3Cu(Cl_{(1-x)}Br_{x})_3}}\xspace}
\newcommand{\DTN}{\ensuremath{\mathrm{NiCl_2\cdot4SC(NH_2)_2}}\xspace}
\newcommand{\DTNX}{\ensuremath{\mathrm{Ni(Cl_{1-x}Br_x)_2\cdot4SC(NH_2)_2}}\xspace}

\journal{Les Comptes Rendus de l'Académie des sciences}

\begin{document}

\begin{frontmatter}



\title{Dirty-boson physics with magnetic insulators}


\author{A. Zheludev}
 \ead{zhelud@ethz.ch}
 \ead[url]{http://www.neutron.ethz.ch/}
 \address{Neutron Scattering and Magnetism, Laboratory for Solid State Physics, ETH Z\"urich, Z\"urich, Switzerland}

\author{T. Roscilde}
 \ead{tommaso.roscilde@ens-lyon.fr}
 \ead[url]{https://sites.google.com/site/roscilde/}
 \address{Laboratoire de Physique, CNRS UMR 5672, Ecole Normale Sup\'erieure de Lyon, Universit\'e de Lyon, 46 All\'ee d'Italie,
Lyon, F-69364, France}

\begin{abstract}
We review recent theoretical and experimental efforts aimed at the investigation of the physics of interacting disordered bosons (so-called {\it dirty bosons}) in the context of quantum magnetism. The physics of dirty bosons is relevant to a wide variety of condensed matter systems, encompassing Helium in porous media, granular superconductors and ultracold atoms in disordered optical potentials, to cite a few. { Nevertheless, the understanding of the transition from a localized, Bose-glass phase to an ordered, superfluid condensate phase still represents a fundamentally open problem. Still to be constructed is also a quantitative description of the highly inhomogeneous and strongly correlated phases connected by the transition. We discuss how disordered magnetic insulators in a strong magnetic field can provide a well controlled realization of the above transition. Combining numerical simulations with experiments on real materials can shed light on some fundamental properties of the critical behavior, such as the scaling of the critical temperature to condensation close to the quantum critical point.}

\end{abstract}


\end{frontmatter}


\section{Introduction}
\label{Introduction}

Quantum magnets have established themselves as an experimental and theoretical testing ground for numerous concepts in  many-body quantum physics. Among these are collective excitations, quantum phase transitions, confinement and asymptotic freedom and geometric frustration, just to name a few  \cite{Sachdev99, Mendels2010}. Disorder and the related spin-impurity effects have been at the center of attention since the early days of quantum and low-dimensional magnetism  \cite{Tippie1981,Fisher1994,Furusaki1995,Eggert1995}. This is hardly surprising, since quantum spin systems are intrinsically more susceptible to disorder than classical ones. The reason is that any spatial randomness of the Hamiltonian modulates the strengths of local quantum fluctuation and often
qualitatively restructures the ground state and excitations.

For a number of reasons, in the few recent years, the study of Hamiltonian disorder in quantum spin systems has experienced a renaissance. First, certain field-induced phase transitions in gapped quantum magnets have been recognized as examples of Bose-Einstein condensation (BEC) of  magnetic quasi-particles (triplons)  \cite{Batyev1984,Giamarchi1999} (for a comprehensive review see  \cite{Giamarchi2008}). The problem of BEC in disorder has long been a very important topic \cite{Giamarchi1988, Fisher1989} that can now be addressed experimentally in spin systems. Second, a breakthrough in materials came with the discovery of numerous ``clean'' realizations of quantum spin models among organic transition metal halogenides  \cite{Patyal1990,Stone2001,Manaka1997,Fujisawa2003}. The magnetic energy scales in such compounds make them much more amenable to experimental investigation than the traditional oxide systems. Moreover, as will be discussed below, reliable ways of introducing controlled disorder into such materials have been developed. From the theoretical point of view, the covalently bonded organic halogenides are easier to describe by simple spin Hamiltonians than the transition metal oxides materials. Last but not least is the recent progress in numerical \cite{Sandvikrev} and experimental \cite{Willis2009, Jaime2010} methods, which have pushed the study of these materials to new levels.

We are  presently in the middle of a new surge in the study of disordered quantum magnets. This is a good time to review some recent achievements, but also outline the outstanding challenges in this exciting and rapidly developing field. In this review we will particularly focus on the effect of disorder in spin-gap magnets exhibiting a magnetic BEC transition, and on the link between the physics of these systems and that of interacting bosons in the presence of disorder. It is worthwhile mentioning that recent progress has been made on other remarkable phenomena exhibited by disordered quantum magnets, such as the random-singlet phase  \cite{Shiroka2011,Thede2012}, the order-by-disorder mechanism induced by impurities in spin-gap compounds  \cite{Azumaetal97,Bobroffetal09,Uchiyama1999}, effect of disorder on excitations  \cite{Huevonen2012-2,Nafradi2013} and on transport properties  \cite{Hess2003,Hlubek2010} to cite only a few.

The structure of this paper is as follows. Sec. \ref{s.bosons} reviews the mapping from a spin-gap system in a magnetic field to a diluted gas of bosons, and the phenomenon of magnetic BEC; Sec. \ref{s.boseglass} introduces the concept of magnetic Bose glass, and Sec.  \ref{s.transition} reviews the known aspects and open questions concerning the Bose-glass/superfluid (or dirty-bosons) transition; Sec. \ref{s.exp} reviews the recent experimental developments in the study of the Bose glass and of the dirty-boson transition in the context of doped halogenides, and Sec. \ref{s.conclusions} discusses conclusions and future perspectives.

\section{Lattice Bose gases from quantum magnets}
\label{s.bosons}

 The connection between spin physics and boson physics has been exploited since the early days of quantum mechanics  \cite{Bloch30}, and it allows to represent the elementary excitations of ordered spin systems as bosonic spin waves. Nonetheless this connection goes beyond the study of ordered phases of magnetic materials, and it remains valid and extremely insightful also in quantum paramagnetic phases, as well as at quantum phase transitions. A particularly prominent example - which will be the focus of the following sections - is provided by magnets in an applied magnetic field (say along the $z$ direction). For spins of length $S$, regardless the ordered or disordered nature of the ground state in a field, one can always represent the magnetization along the field direction, $S_i^z$, at the $i$-th lattice site in terms of the number of bosonic quasiparticles, $n_i = b_i^{\dagger} b_i = S + l ~S_i^z$ (with $l = \pm 1$), constrained by the condition $0 \leq n_i \leq 2S$. Here $b_i$, $b_i^{\dagger}$ are bosonic operators, satisfying the commutation relation $[b_i, b_j^{\dagger}]=\delta_{ij}$. The algebra of commutators between the spin components can then be completed via {\it e.g.} the Holstein-Primakoff transformation  \cite{HolsteinP40}, which for $l = 1$ reads $S_i^- = \sqrt{2S-n_i} ~b_i$, $S_i^+ = b_i^{\dagger} \sqrt{2S-n_i}$. The expressions of $S_i^+$ and  $S_i^-$ are exchanged for $l = -1$.

 This transformation therefore maps formally {\it any} quantum spin system to a system of bosonic quasiparticles.
We would like to stress that, if the system does not exhibit long-range magnetic order along the $z$ axis, $n_i$ does {\it not} correspond to the local density of elementary excitations in the system - as in spin-wave theory. Therefore the bosons might {\it not} correspond to spin-wave quanta (or magnons), but rather magnetic quasiparticles with a finite equilibrium density which can be non-zero even at $T=0$. The density is indeed controlled by the magnetic field $H$, which acts as a chemical potential through the Zeeman Hamiltonian term
\begin{equation}
{\cal H}_{\rm Zeeman} = - g \mu_B H \sum_i S_i^z = - \mu \sum_i n_i + {\rm const.}
\end{equation}
where $\mu = l g \mu_B H$, $g$ is the atomic gyromagnetic factor, and $\mu_B$ the Bohr magneton. This situation is generally quite different with respect to spin-wave theory, in which a spin-boson transformation, a linearization of the Hamiltonian and (if necessary) a Bogolyubov transformation transform the spin system to a lattice gas of free bosons with a {\it vanishing} chemical potential. This aspect has raised some confusion in the recent literature  \cite{BunkovV10, BunkovV10-2}.

The point of view of bosonic quasi-particles is certainly most insightful if the particle number is conserved in the bosonic Hamiltonian - as this allows to draw a full analogy between the magnetic phases and the phases of lattice-boson Hamiltonians. Particle number conservation requires an axial rotation symmetry around the field axis (U(1) symmetry) to be present in the Hamiltonian. Such a continuous symmetry is never exact in magnetic compounds due to the crystalline field (see discussion in Sec.~\ref{s.materials}), but small deviations become only relevant in the very close neighborhood of phase transitions.

\subsection{Magnetic Bose-Einstein condensation at equilibrium}
\label{s.mBEC}

 The exact correspondence between axially symmetric spin Hamiltonians in a field and lattice boson Hamiltonians in the grand-canonical ensemble allows to investigate a most fundamental phenomenon of lattice bosons using magnetic insulators: the quantum phase transition from a superfluid BEC \footnote{The superfluid nature of the condensate phase realized by the Hamiltonians of magnetic insulators is well established theoretically, but in practice current experiments cannot probe superfluidity - which would imply probing persistent spin currents. At a more fundamental level, the absence of a strictly exact particle-number conservation in the experiments could lead to a rapid decay of supercurrents, so that some fundamental aspects of superfluidity would not be observable.} to an insulator  \cite{Sachdev99, Giamarchi2008}.
The simplest example of such a transition is obtained for the $S=1/2$ XXZ Hamiltonian in a magnetic field
\begin{equation}
{\cal H} = J\sum_{\langle ij \rangle} \left( S_i^x S_j^x + S_i^y S_j^y + \lambda S_i^z S_j^z  \right)- g\mu_B H \sum_i S_i^z~
\label{e.XXZ}
\end{equation}
where $\langle ij \rangle$ represents a pair of nearest neighboring sites on the lattice, and $J>0$.
We will consider for simplicity the case in which this Hamiltonian is defined on a bipartite lattice (namely in the absence of frustration); in this case the sign of the coupling for the $x$ and $y$ spin components can be changed arbitrarily via a canonical transformation $S_i^{x(y)} \to (-1)^i S_i^{x(y)}$.

\begin{figure}
\begin{center}
\mbox{\includegraphics[width=9cm]{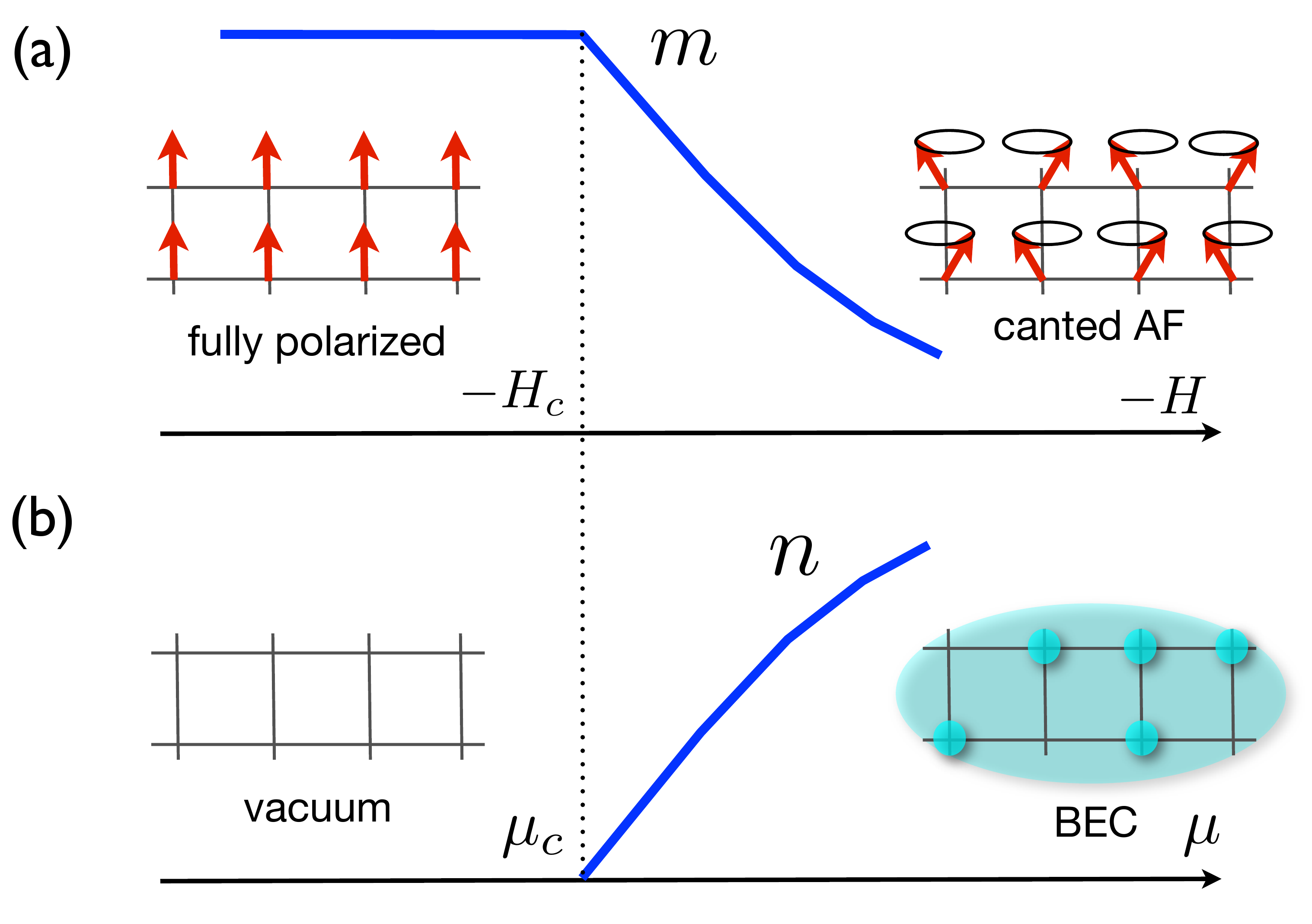}~~~~~
\includegraphics[width=7cm]{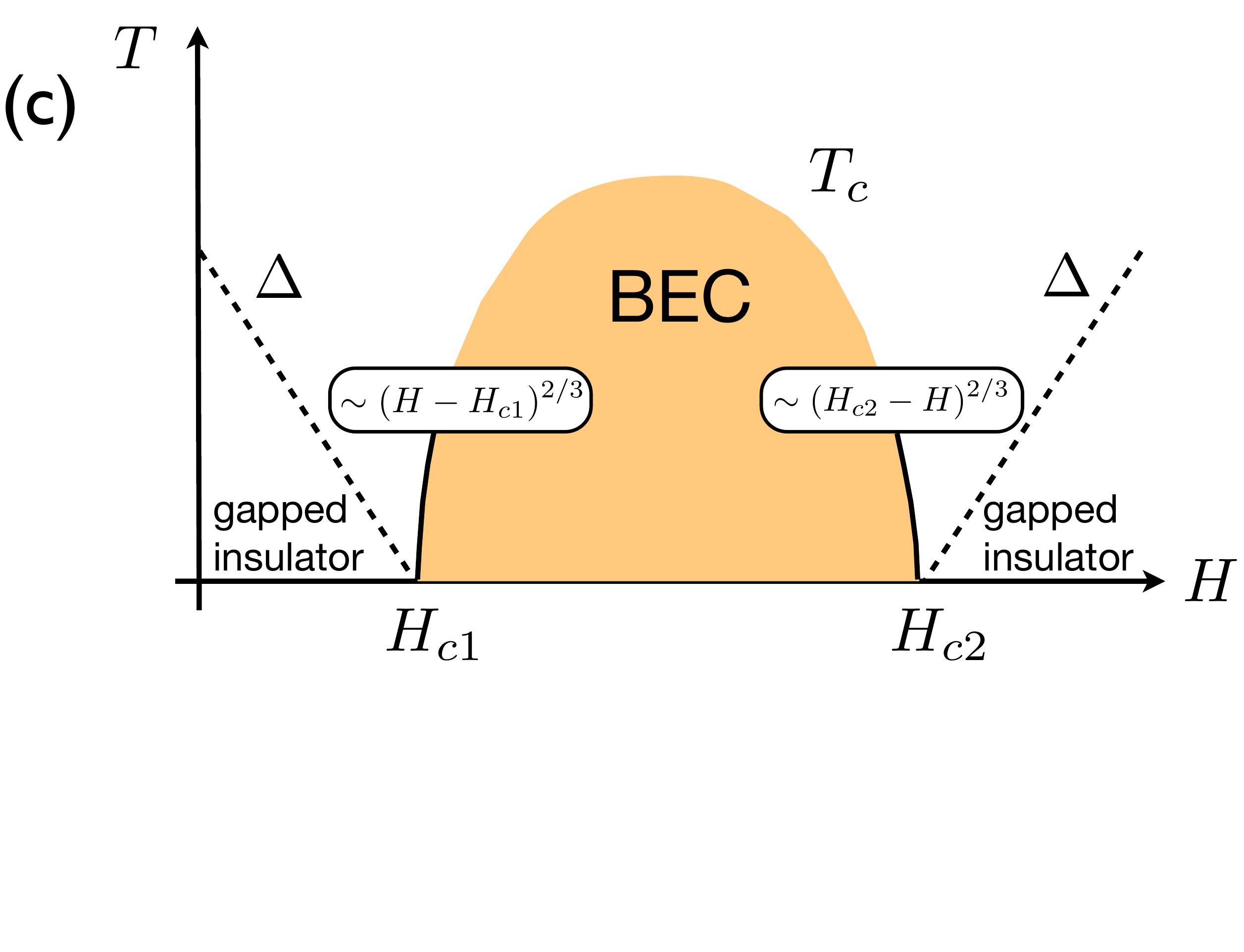}}
\caption{Sketch of the magnetic Bose-condensation transition for (a) the XXZ model, Eq.~\eqref{e.XXZ} and (b) its bosonic analog, Eq.~\eqref{e.BH}; here $m = \sum_i \langle S_i^z \rangle /N$ represents the field-induced magnetization of the system, and $n = S -m$ the density of bosonic quasiparticles.
(c) Generic temperature-field phase diagram.}
\label{f.sketch}
\end{center}
\end{figure}

As first noticed by Matsubara and Matsuda  \cite{MatsubaraM56}, in the case of $S=1/2$ the spin-boson transformation simplifies greatly when considering {\it hardcore} boson operators $\tilde{b}_i$ and $\tilde{b}_i^{\dagger}$, which anticommute on-site $\{ \tilde{b}_i, \tilde{b}_i^{\dagger} \} = 1$, $\{ \tilde{b}_i, \tilde{b}_i \} =  \{ \tilde{b}_i^{\dagger}, \tilde{b}_i^{\dagger} \} = 0$ (but they still commute offsite).  In this case, choosing $l = -1$, one has $\tilde{n}_i = \tilde{b}_i^{\dagger} \tilde{b}_i = S - S_i^z$, $S_i^+ = \tilde{b}_i$, $S_i^- = \tilde{b}^{\dagger}_i$, and the spin Hamiltonian takes the form of an extended Bose-Hubbard model  \cite{MatsubaraM56}
\begin{equation}
{\cal H} = -\frac{J}{2} \sum_{\langle ij \rangle} \left( \tilde{b}_i^{\dagger} \tilde{b}_j + {\rm h.c.} \right) + \lambda J \sum_{\langle ij \rangle} \tilde{n}_i \tilde{n}_j -  \left( \frac{\lambda Jz}{2}  -  g\mu_BH \right) \sum_i \tilde{n}_i  + {\rm const.}~
\label{e.BH}
\end{equation}
where $z$ is the coordination number.
For $\lambda > -1$, the spin Hamiltonian is known to exhibit a continuous transition from a canted antiferromagnetic state at small or intermediate fields, to a fully saturated phase at $H > H_c =  \frac{z}{2 g \mu_B }(\lambda + 1) J$ for a hypercubic lattice in $d = z/2$ spatial dimensions.
The transition is sketched in Fig.~\ref{f.sketch}.
After mapping onto bosonic operators with $l=-1$, the canted antiferromagnetic phase is immediately identified with a phase with finite, incommensurate density and with off-diagonal long-range order, $\langle \tilde{b}_i^{\dagger} \tilde{b}_j \rangle = \langle S_i^- S_j^+ \rangle = (1/2) ~\langle S_i^x S_j^x + S_i^y S_j^y \rangle \to (-1)^{i-j} const. \neq  0$   when $|i-j|\to \infty$, corresponding to condensation (in one spatial dimension, the system exhibits quasi long-range order with algebraically decaying correlations). On the other hand, the fully saturated phase with exact ground state
$\otimes_{i=1}^N |S_i^z = 1/2\rangle$ corresponds to the most trivial bosonic insulator, namely the vacuum $\otimes_{i=1}^N |n_i =0\rangle$, with a zero-ranged phase correlation function and a finite spin gap $\Delta  = g\mu_B (H - H_c)$. The critical field corresponds to the condition of the chemical potential $\mu = \frac{\lambda Jz}{2}  -  g\mu_BH $ touching the lower band edge for the lattice bosons, $\mu = - zJ/2$.

The magnetic Bose-condensation transition represents one of the best understood quantum phase transitions, belonging to the universality class of the diluted Bose gas  \cite{Sachdev99}. A path-integral representation of the partition function for the Hamiltonian Eq.~\eqref{e.BH} (in which the hardcore constraint can be relaxed in favor of an arbitrarily strong repulsive interaction), supplemented with a Hubbard-Stratonovich transformation, maps the system to a quantum complex $\phi^4$ theory  \cite{Fisher1989}, which, in any dimension $d$, is known to possess a ``generic" quantum phase transition with dynamical critical exponent $z=2$, belonging to the mean-field universality class. Above the upper critical dimension, $d>d_c=2$, the fixed point corresponding to this transition is a Gaussian model, for which the interaction between quasiparticles can be treated perturbatively and turns out to be irrelevant - more precisely it is {\it dangerously} irrelevant, affecting the scaling close to the transition, and leading to a breakdown of the quantum hyperscaling relation $\nu (d+z) = 2- \alpha$ (where $\alpha$ is the free-energy critical exponent and $\nu$ is the correlation length critical exponent).

In the case of a $d=3$ system, condensation in the BEC phase persists up to a critical temperature $T_c$ whose scaling with the distance to the critical point can be obtained via Popov's theory for the diluted Bose gas  \cite{Popov87, Giamarchi1999}, to give
$T_c \sim |\mu-\mu_c|^{\phi} \sim |H-H_c|^{\phi}$ with $\phi=2/3$ - as sketched in Fig.~\ref{f.sketch}(c).

\subsection{Realizations of magnetic Bose condensation}
 As discussed above, the saturation transition of an axially symmetric antiferromagnet (as well as of a ferromagnet with easy-plane anisotropy) represents the simplest realization of a magnetic BEC transition. This transition has the experimentally inconvenient feature that extremely high (or even prohibitively high) magnetic fields might be required in order to saturate the magnetization real in magnetic insulators. A convenient alternative to the     above scenario for magnetic BEC is offered by magnetic compounds which can be thought of realizing an insulating state of bosonic quasiparticles already in {\it zero} applied field; applying a field drives the system towards an insulator-to-superfluid transition at a lower critical field ($H_{c1}$), followed by a second superfluid-to-insulator transition at an upper critical field ($H_{c2}$) corresponding to the above-mentioned saturation (see Fig.~\ref{f.sketch}(c)).
 Experimentally relevant examples of such systems are: a) weakly coupled antiferromagnetic dimers; b) integer-$S$ spins with a strong single-ion anisotropy; c) weakly coupled Haldane chains.

 {\it Weakly coupled dimers.} In the case of weakly coupled dimers, {\it e.g.} for $S=1/2$, the building block - a spin dimer with antiferromagnetic coupling $J_d$ - is in a singlet ground state $|s\rangle = \left(|\uparrow\downarrow \rangle - |\downarrow \uparrow\rangle\right)/\sqrt{2}$, separated by a gap $\Delta=J_d$ from three triplets. Under application of a magnetic field the triplet aligned with the field, $|t_1\rangle = |\uparrow \uparrow \rangle$, is brought towards degeneracy with the singlet, while the other two triplets remain well separated in energy. In a coupled dimer system, the coupling $J$ between dimers, being much weaker than the dimer gap, $J \ll \Delta = J_d$, leaves the energy of the upper triplets essentially unaffected, so that they can be eliminated as far as the low-energy properties of the system are concerned. As a consequence each dimer can be approximated as a two-state system, $|s\rangle$ and $|t_1\rangle$, analog to a pseudo-spin $S=1/2$. The effective Hamiltonian for coupled dimers in a magnetic field in terms of pseudo-spin variables takes precisely the form of an XXZ Hamiltonian as in Eq.~\eqref{e.XXZ}  \cite{TachikiY70,Mila98}, and therefore the same BEC transition is expected.

 {\it Anisotropic integer-$S$ systems.} A very similar mapping can be obtained in the case of integer-$S$ spins with a strong single-ion anisotropy $D(S_i^z)^2$. The anisotropy energy is minimized by the $|m_s=0\rangle$ state, where $|m_s \rangle$ indicates an eigenstate of the $S_i^z$ operator, and a gap $\Delta=D$ separates this state from the states with $m_s \neq 0$. The application of a magnetic field brings the state with $m_S=S$ to degeneracy with the $m_S=0$ ground state, keeping the other states higher up in energy by at least an energy $D$. If the spin-spin coupling $J$ is much weaker than $D$, truncating the Hilbert space to $|m_s=0\rangle$ and $|m_s=S\rangle$ leads again to a $S=1/2$ pseudo-spin system governed by the XXZ Hamiltonian of Eq.~\eqref{e.XXZ}  \cite{TsunetoM71}.

 {\it Haldane chains.} As for weakly coupled Haldane chains in a field, the connection to the physics of the diluted Bose gas is more involved, and it requires a mapping of the spin Hamiltonian to a non-linear $\sigma$-model, as first described in  \cite{Affleck90, Affleck91, Mitra94} (see also  \cite{Zheludev2004} for a critical discussion). The BEC nature of the field-induced transition has been verified numerically  \cite{Sorensen1993}.

\begin{figure}
\begin{center}
\mbox{\null\hspace*{-.8cm}
\includegraphics[width=9cm]{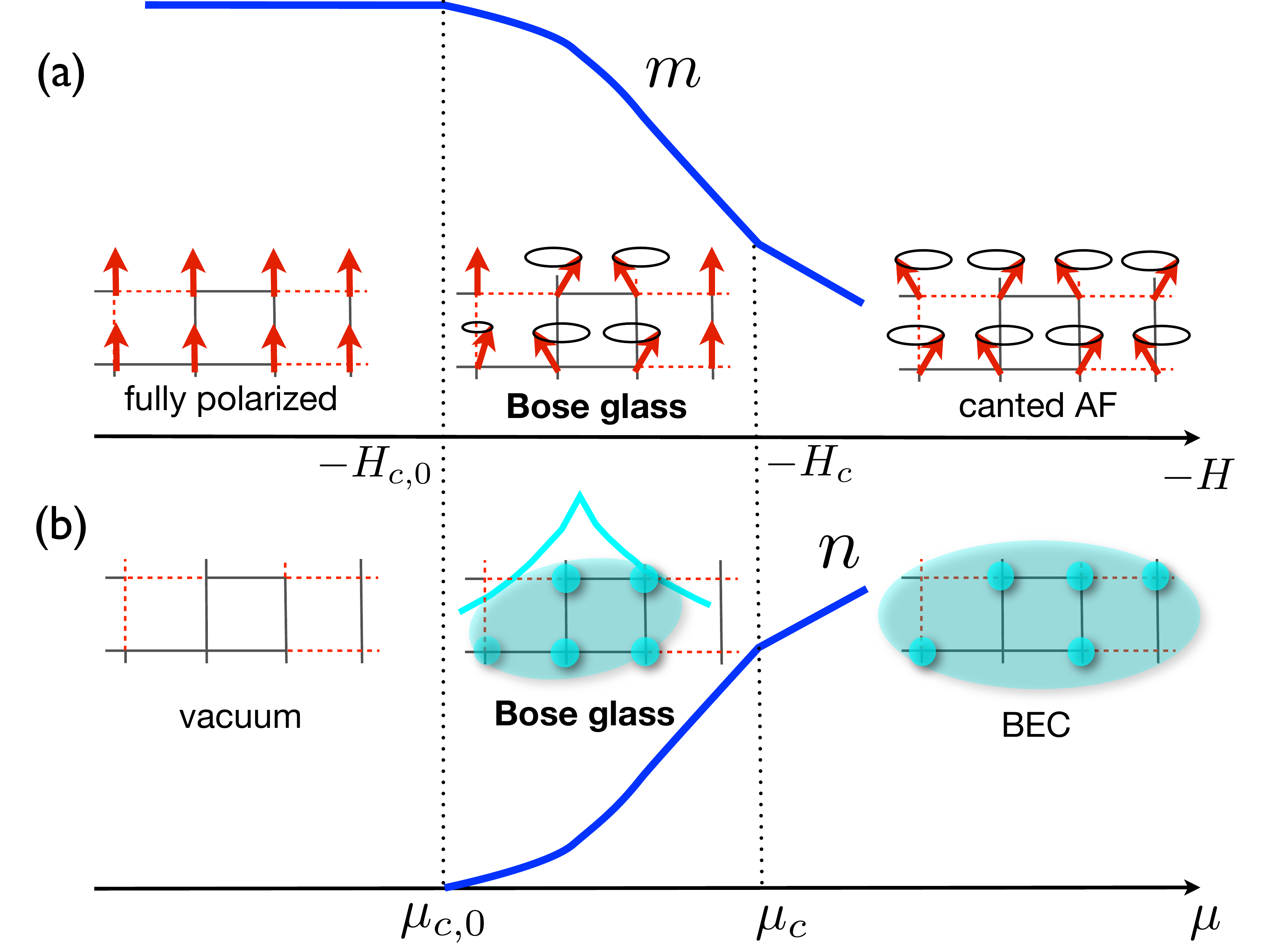}~~~~~
\includegraphics[width=8cm]{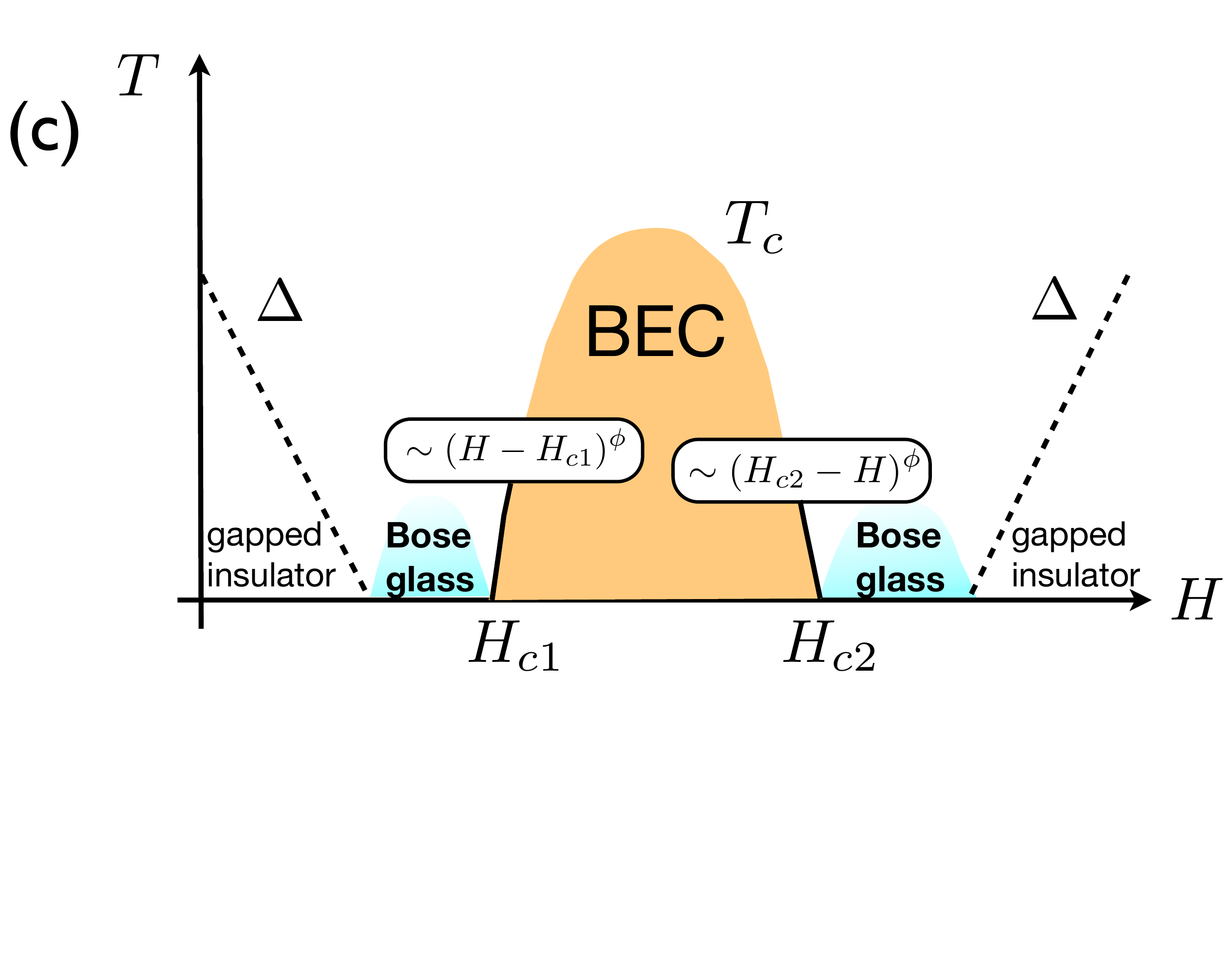}}
\caption{Sketch of the magnetic Bose-condensation transition in the presence of disorder. (a) Sketch of the phases of the disordered XXZ model; (b) corresponding phases for the bosonic analog; (c) generic temperature-field phase diagram.}
\label{f.sketch-dis}
\end{center}
\end{figure}


\section{The magnetic Bose Glass}
\label{s.boseglass}
\subsection{Basic concepts}

 Disorder can have a dramatic impact on the physics of the diluted Bose gas, and hence on magnetic BEC. As we will further discuss in Sec.~\ref{s.materials}, magnetic insulators offer various ways of disordering the Hamiltonian Eq.~\eqref{e.XXZ} and its bosonic analog Eq.~\eqref{e.BH}, as for instance via the appearance of random couplings $J$, or via site dilution of the magnetic lattice.
 From the point of view of bosonic quasiparticles, disorder effects are intimately related to the drastic change of the (low-energy) single-particle eigenstates in the presence of disorder, from extended propagating states to exponentially localized states \`a la Anderson  \cite{KramerMK93}. Condensation and Anderson localization are two conflicting concepts in the presence of weak, albeit finite repulsive interactions. Indeed, upon increasing the particle number along with the system size, finite repulsion prevents bosons to condense in a single localized state: by definition, the support of such a state  does not scale with the size of the system, and hence the (mean-field) interaction energy of a localized condensate would grow like the square of the particle number. An extensive ground-state energy is instead obtained by fragmenting the condensate over an extensive number of spatially separated localized states. In the dilute limit the system is therefore in a non-condensed, Anderson insulating phase: the {\it Bose glass}  \cite{Fisher1989}. At variance with the condensate phase, in the Bose glass the order-parameter correlation function decays exponentially over a distance related to the typical spatial extent of the above-mentioned localized states. Yet the Bose glass is also substantially different from the gapped insulating phase of the bosonic model. Indeed the single-particle spectrum in the presence of disorder has generically a finite density of states down to the ground state energy, due to the existence of an exponentially decaying Lifshitz tail associated with excitations localized in rare, locally uniform regions. This implies that the Bose glass is a gapless, compressible bosonic insulator. When increasing the chemical potential and closing the gap, the system is therefore driven first to a Bose glass phase for low density, while condensation can be achieved only at higher densities (Fig,~\ref{f.sketch-dis}), as discussed in the next section.

 Mapping the bosons back to spins, what is the magnetic analog of the Bose glass? Following the picture of the XXZ model (Fig.~\ref{f.sketch-dis}(a)), in the magnetic Bose glass a finite fraction of the system is characterized by the appearance of spatially separated, localized domains in which the spins can fluctuate away from the fully polarized state, exhibiting short-range correlations between their transverse components - nonetheless different domains are not magnetically correlated. For a generic magnetization $m = \sum_i \langle S_i^z \rangle /N$ below the saturation value, a finite fraction of these domains (and hence an extensive fraction of the whole system) exhibits a nonzero response to a variation of the magnetic field, leading to a finite global magnetic susceptibility (the analog of the compressibility) down to zero temperature.

\subsection{A minimal model for the Bose glass}

A very simple (yet quantitatively predictive) model for the Bose glass phase has been introduced in Refs.~ \cite{Roscilde06, Yu2012}, and consists of modeling the fluctuating domains as two-level systems with a size-dependent local gap ({\it local gap model}).  The lower level corresponds to the local vacuum of quasiparticles, while the upper level corresponds to the presence of a localized bosonic quasiparticle. A fluctuating region is a rare, locally uniform region, {\it e.g.} devoid of impurities, or uniformly rich of impurities, if disorder is introduced via chemical doping (as it will be the case in the magnetic realizations of Bose glass described in Sec.~\ref{s.materials}). In the case of short-range correlated disorder, we can assume that rare regions of size $N$ occupy a fraction $f_N \approx A \exp(-x_0 N)$ of the whole system (where $x_0$ is a parameter related to the disorder strength, {\it e.g.} $x_0 = |\log (1-x)|$ where $x$ is the impurity concentration if the rare region is devoid of impurities). We further assume that a rare region would exhibit long-range order (namely BEC) if scaled to size $N\to \infty$ when the chemical potential exceeds the critical value $\mu_{c,0}$ for the clean system. Hence we can expect rare regions to exhibit a local gap to excitations which scales as $\Delta_N \approx \delta_0 ~\theta(-\delta_0) +  c/N$ where $\delta_0 = \mu - \mu_{c,0}$ is the distance to the clean system critical point, $\theta$ is the Heaviside function, and $c$ is a constant.  The gap vanishes as $1/N$ in the BEC phase of the clean system, as the lowest energy excitation of the Anderson's tower  \cite{Andersonbook} in a phase with spontaneous symmetry breaking.
The density of states of the rare region ensemble is then expressed as
\begin{equation}
g(\epsilon) = \sum_{N=1}^{\infty} \delta(\epsilon- \Delta_N) f_N
\end{equation}
and it exhibits the following behaviors:
\begin{itemize}
\item for $\delta_0 > 0$, namely in the gapped insulator phase, it vanishes for $0 \leq \epsilon \leq \delta_0$;
\item at the clean critical point $\delta_0 = 0$ it exhibits a Lifshitz tail $g \sim \exp(-x_0 c/\epsilon)$;
\item in the Bose glass phase for $\delta_0 > 0$ it goes to a finite value at zero energy, $g(0) \sim \exp(-x_0 c/\delta_0)$.
\end{itemize}
The quasiparticle density in the ground state of the rare-region ensemble is directly proportional to the density of states, $n \sim \exp(-x_0 c/\delta_0)$, and hence the system exhibits a finite ground-state compressibility $\kappa = dn/d\mu \sim  \exp(-x_0 c/\delta_0)$ in the Bose glass phase. The compressibility as well as the gap vanish at the special value $\delta_0=0$, realizing a so-called {\it Mott glass} point.

\section{The Bose-glass/BEC transition}
\label{s.transition}

 In the presence of disorder, quasi-particle interactions have a disruptive effect on the condensate in the regime of very low densities, as already discussed in the previous section. On the other hand, for larger densities repulsive interactions reduce the effect of disorder via a screening mechanism, and they can ultimately restore condensation upon increasing the chemical potential. In the magnetic language, this corresponds to recovering long-range order transverse to the field when the magnetization deviates sufficiently from the value of the disordered, gapped phase.
 Hence the system undergoes a quantum phase transition from Bose glass to superfluid BEC (the so-called the {\it dirty-boson transition}), driven by the chemical potential / magnetic field. This insulator/superfluid quantum phase transition is fundamentally different from the one of the clean system: 1) it takes place at densities arbitrarily far from the diluted regime, and it is fundamentally driven by interactions; 2) it connects highly inhomogeneous phases, namely a collection of localized domains fluctuating in an uncorrelated manner (the Bose glass) to a system of localized domains which become correlated via the coherent tunneling of particles along weak links (the inhomogeneous BEC close to the transition). Its theoretical description represents one of the most important open problems in the theory of critical phenomena, as we will discuss in the following.

\subsection{$d=1$. Weak vs. strong disorder}

 The study of the dirty-boson transition has been pioneered in Refs.~ \cite{Giamarchi1987, Giamarchi1988} for the case of one-dimensional bosons - for which the BEC phase is actually a quasi-condensate phase, exhibiting a finite superfluid response. The long-wavelength effective theory (Luttinger liquid theory) for bosons in a disordered potential has been investigated via the renormalization group, treating the disorder perturbatively (originally to first order, but more recently the analysis has been extended to second order  \cite{Ristivojevicetal12}, confirming the first-order results). The main outcome of this analysis is that the phase transition has a Kosterlitz-Thouless (KT) nature, and it is characterized by a universal jump of the Luttinger liquid parameter $K$ governing the algebraic decay of the correlation function $\langle b_i^{\dagger} b_j \rangle \sim |r_i-r_j|^{-1/{(2K)}}$ (here expressed for lattice systems of interest in this paper): one finds that $K = K_c = 3/2$ at the transition point  (while $K_c=2$ for the superfluid-insulator transition in a clean lattice system with commensurate filling), and that the superfluid quasi-condensate is characterized by $K > K_c$. While the above result applies to the regime of {\it weak} disorder, it has been confirmed by the analysis of Ref.~ \cite{Kashurnikovetal96} based on Popov's effective hydrodynamic theory for the interacting Bose gas  \cite{Popov87}, relying on the weaker assumption of self-averaging of the compressibility and superfluid stiffness. On the other hand, a different picture emerges from real-space renormalization group studies of quantum rotor lattice models with {\it strong} disorder, indicating a KT transition with a {\it non-}universal critical value $K_c$ dependent on disorder  \cite{Altmanetal08, Altmanetal10}.
  The reconciliation between the weak- and strong-disorder pictures appears quite challenging, as each picture is supposed to be quantitatively accurate in its own range of applicability. Recently Ref.~ \cite{Polletetal13} has provided quantum Monte Carlo results consistent with a universal value $K_c=3/2$ at the transition in a \emph{strongly} disordered link-current model; the Luttinger parameter estimate turns out to be strongly size-dependent, and close to the phase transition it converges logarithmically with system size to its asymptotic value, indicating that prohibitively large samples might be necessary to observe the correct physics.

  A similar controversy applies to the case of the dirty-boson transition in a system at commensurate filling, namely for the transition driven {\it e.g.} by the strength of the interaction at fixed, integer  density, corresponding to a fixed magnetization. In this case the transition connects a superfluid with an incompressible, gapless Mott glass  \cite{Giamarchietal01, ProkofevS04}. For this case QMC results on the disordered link-current model  \cite{Balabanyanetal05} point to a conventional KT transition with $K_c=2$, unaltered with respect to the clean case, in agreement with an analysis based on Popov's hydrodynamic theory. The latter results contradict a previous real-space RG analysis indicating a disorder-dependent $K_c$ value  \cite{Altmanetal04}. A more recent QMC study of disordered quantum rotors  \cite{HrahshehV12}, interpolating between the two regimes of weak and strong disorder,  shows that $K_c$ might indeed become disorder-dependent at sufficiently strong disorder.

\subsection{$d>1$. Scaling theory and its consequences}

 The study of the dirty-boson transition in dimensions $d>1$ it has been pioneered by Ref.~ \cite{Fisher1989}.  The latter reference has obtained an effective replicated $\phi^4$ action for the description of the critical behavior of the disordered Bose gas, but it has shown that a perturbative renormalization group (RG) approach to the transition fails, as the disorder grows with the RG flow (see also Refs.~ \cite{WeichmanK89, WeichmanM08} for successive studies). Yet a few simple arguments led the authors of Ref.~ \cite{Fisher1989} to formulate a phenomenological scaling theory leading to some fundamental, yet highty debated predictions. In the presence of disorder, the Harris criterion $\nu > 2/d$  \cite{Chayesetal86} is satisfied by the mean-field exponent $\nu = 1/2$ for $d>4$, so that the upper critical dimension is $d_c\geq 4$ (Ref.~ \cite{Fisher1989} argues that possibly $d_c=\infty$). All physical dimensions $d \leq 3$ are below the upper critical one, and conventional scaling can be assumed. Hence it is in principle legitimate to assume that, beside the temperature, all perturbations at the quantum critical point are strictly irrelevant, leading to the simplest scaling form for the singular part of the free energy density
 \begin{equation}
 f_s(T,\delta) \approx |\delta|^{2-\alpha}~ F\left(\frac{T}{|\delta|^{\nu z}}\right) = |\delta|^{2-\alpha}~ Y\left(\frac{\xi^z}{\beta}\right)
 \label{e.fs}
 \end{equation}
 where $\delta = \mu - \mu_c$ represents the distance to the critical point, and $\xi \sim |\delta|^{-\nu}$ is the correlation length. Such a scaling form implies quantum hyperscaling, $2-\alpha = \nu(d+z)$.
 Introducing an imaginary-time twist of the Bose field,  which becomes infinitesimal in the $T\to 0$, $\beta \to \infty$ limit, namely a phase shift $\Delta\theta/\beta$, amounts to a shift of the chemical potential, $\mu \to \mu -i ~\Delta\theta/\beta$. The corresponding variation of the (total) free energy density reads
 \begin{equation}
 \Delta f = i\rho \left(\frac{\Delta \theta}{\beta}\right)+ \frac{1}{2} \kappa \left(\frac{\Delta \theta}{\beta}\right)^2 + ...
 \end{equation}
 where $\rho =  -\partial f/\partial \mu$ is the particle density and $\kappa = \partial \rho/\partial \mu$ is the compressibility. Ignoring arbitrarily the first term of the above expansion, and assuming (somewhat arbitrarily) that the second-order term of the total free energy density should scale in the same way as the singular part, Eq.~\eqref{e.fs}, one can conclude that $Y(x) \sim x^2$ and that $\kappa \sim \delta^{\nu(d-z)}$. Now, given that the Bose-glass/superfluid transition connects two compressible phases, one can argue that $\kappa$  remains finite across the transition, leading to the conclusion that $z=d$ for $d<d_c$.

  A further consequence of the scaling assumption, Eq.~\eqref{e.fs}, concerns the onset of the critical temperature in the BEC phase. As Eq.~\eqref{e.fs} is supposed to hold for finite temperatures close to the quantum critical point, the line of phase transitions at $T_c = T_c(\delta)$ corresponds to a singularity $y_c$ of $F(y)$, leading to the scaling $T_c = y_c |\delta|^{\nu z}$. Combining the prediction $z=d$ with the Harris criterion leads to the inequality $\nu z > 2$, so that $T_c \sim |\delta|^{\phi}$ with $\phi>2$, to be contrasted with $\phi=2/3$ for the clean system.

\subsection{Numerical results and controversies with the scaling theory}

The two main predictions coming from the scaling theory of the Bose-glass/superfluid transition, namely $z=d$ and $\phi>2$, have been both strongly debated in the more recent literature.

\bigskip
\noindent {\it $z=d$.}

\noindent Some quantum Monte Carlo (QMC) studies in $d=2$, performed on hardcore bosons  \cite{Priyadarsheeetal06} and on the link-current model  \cite{MeierW12}, have found that $z=d=2$ is not verified at the Bose-glass/superfluid transition (these studies find $z \approx 1.4$ and $z = 1.80(5)$ respectively); in particular Ref.~ \cite{MeierW12} points out that the numerical estimate of $z$ might be strongly size dependent. These results have prompted a critical reconsideration of the scaling theory. Refs.~ \cite{WeichmanM07,WeichmanM08} (including one of the authors of the scaling theory) have pointed out the above-mentioned arbitrariness of the assumptions leading to the compressibility scaling $\kappa \sim \delta^{\nu(d-z)}$, and they have proposed that $z$ and $d$ be actually independent exponent. In particular the main weakness of the argument leading to $z=d$ according to Refs.~ \cite{WeichmanM07,WeichmanM08} is the assumption that the free-energy-density variation upon imaginary-time twist obeys the same scaling as that of the \emph{singular} free energy density. This assumption is indeed contradictory with the fact that the singular compressibility $\kappa_s = -\partial^2 f_s/\partial \mu^2 \sim |\delta|^{-\alpha}$ vanishes at the transition if hyperscaling is satisfied, as $\alpha = 2-\nu(d+z) < 0$ according to the Harris criterion (whatever the value of $z>0$). Therefore a finite compressibility across the transition is dominated by the analytical part of the free energy density, whose scaling is not necessarily captured by Eq.~\eqref{e.fs}.

 On the other hand further QMC studies  \cite{ProkofevS04, Linetal12, Soyleretal11} on softcore bosons in $d=2$ observe scaling of the numerical results consistent with $z=2$; scaling properties consistent with $z=3$ are also observed for $d=3$ link-current simulations  \cite{HitchcockS06}, as well as for $d=3$ disordered spin Hamiltonians modeling the field-induced transition of one of the compounds discussed in Section \ref{s.materials}  \cite{Yuetal12-2}. These results suggest that $z=d$ might still be valid, and that the derivation of Ref.~ \cite{Fisher1989} gives coincidentally the correct result. Indeed the critical scaling $\kappa \sim |\delta|^{\nu(d- z)}$ is observed at the diluted Bose gas transition in the clean case in dimensions $d\leq 2$ (namely below the upper critical dimension), it is observed at the superfluid-Mott insulator transition at commensurate filling \cite{Fisher1989}, as well as at the dirty-boson transition in $d=1$  \cite{Fisher1989}. To corroborate this result, a very simple derivation of the compressibility scaling, alternative to that of Ref.~ \cite{Fisher1989}, can be obtained  \cite{Yuetal12-2} based on the following consideration. At $T=0$, $\kappa \approx (\Delta_{\rm ph} L^d)^{-1}$, where $\Delta_{\rm ph}$ is the gap for a particle-hole excitation, $\Delta_{\rm ph} = E(N+1) + E(N-1) - 2E(N)$ with $E(N)$ the ground state energy for $N$ particles, and $L$ is the linear size of the system. Given that by definition $\Delta_{\rm ph} \sim L^{-z}$ at the quantum critical point, one readily concludes that $\kappa \sim L^{-(d-z)} \sim \xi^{-(d-z)} \sim |\delta|^{-\nu (d-z)}$.

\bigskip
\noindent  {\it $\phi>2$.}

\begin{figure}
\begin{center}
\includegraphics[width=8cm]{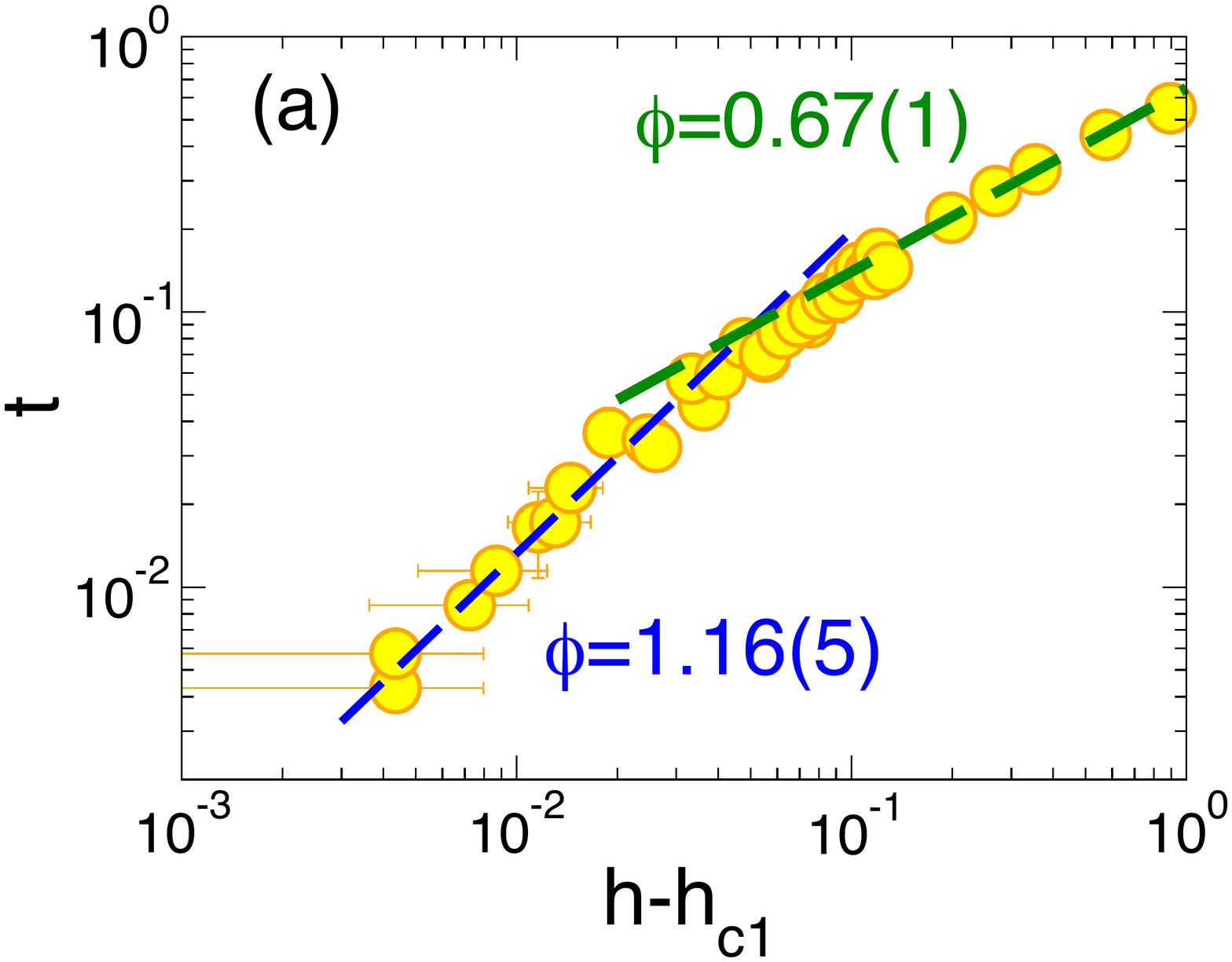}
\includegraphics[width=7.7cm]{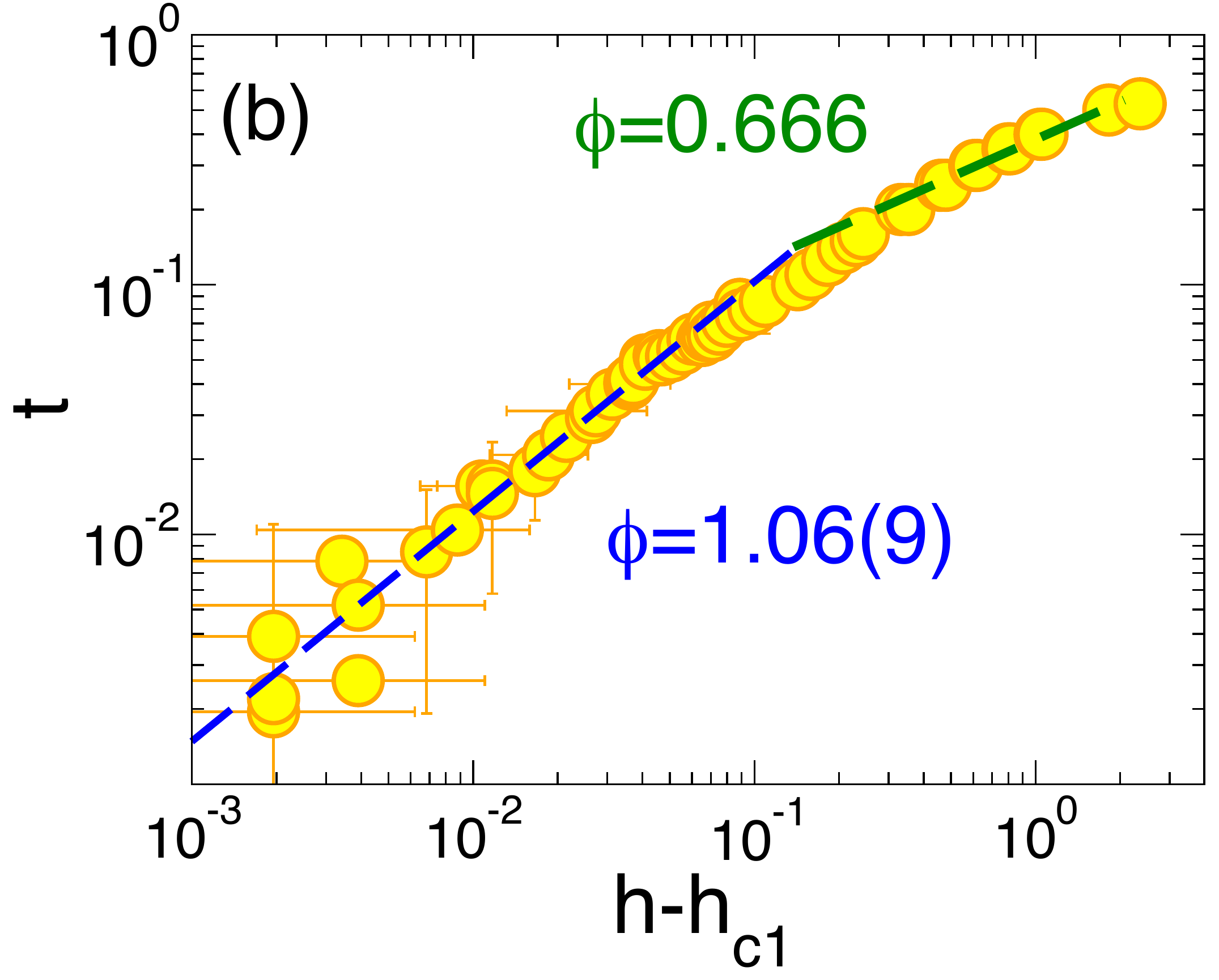}
\caption{Scaling of the critical temperature in disordered $S=1$ antiferromagnets with strong single-ion anisotropy, as obtained from extensive quantum Monte Carlo simulations  \cite{Yu2010epl, Yu2012}: (a) 15\% site dilution; (b) 15\% bond disorder. The axes represent the field strength $h = g \mu_B H/J_c$ and the temperature $t=k_B T/J_c$ in units of the strongest coupling $J_c$ of the Hamiltonian. Dashed lines are power-law fits to the low-temperature and intermediate-temperature regimes. The figure is adapted from Refs.~\cite{Yu2010epl,Yu2012}.}
\label{f.phitheory}
\end{center}
\end{figure}

\noindent  The second prediction of scaling theory, $\phi>2$, is currently emerging as the most controversial. To the best of our knowledge it is not corroborated by any numerical result to date. On the other hand, extensive QMC simulations for disordered $S=1$ antiferromagnets with strong single-ion anisotropy in a magnetic field point at a value $\phi\approx 1.1$  \cite{Yu2010epl, Yu2012}, as shown in Fig.~\ref{f.phitheory}. Further predictions for the temperature dependence of the magnetization and specific heat along the quantum critical trajectory ($\delta=0$, $T\to 0$) stemming from Eq.~\ref{e.fs} disagree with numerics  \cite{Yuetal12-2}.
A phenomenological understanding of all these discrepancies requires to generalize the scaling Ansatz, Eq.~\eqref{e.fs}, in order to include further arguments in the scaling function. An attempt in this direction is given by Ref.~ \cite{Yuetal12-2}, in which alternative scaling forms are proposed featuring unconventional scaling activated only at finite temperature - given that all available numerical results are consistent with conventional scaling at the $T=0$ quantum phase transition.

\bigskip
\noindent  {\it Other critical exponents.}

\noindent A finite-size scaling analysis of quantum Monte Carlo data for the dirty-boson transition in $d=3$ is presented in Refs.~ \cite{HitchcockS06} and  \cite{Yuetal12-2} for the link-current model and the single-ion anisotropic $S=1$ magnets, respectively.  A coherent picture emerges, consistent with $z\approx 3$, $\beta \approx 0.95$, $\nu\approx  0.75$, $\eta\approx -1$.

\section{The quest for a ``quantum simulator" in doped\protect\footnote{ In its strict definition, ``doping'' implies the introduction of free charge into a semiconductor or insulator by means of a chemical substitution. In the present context by ``doping'' we only mean a low-concentration chemical modification of the parent compound.} magnetic insulators}
\label{s.exp}

The above results show that many crucial points of the dirty-boson transition are still an active matter of debate in the theoretical community. The whole theory suffers from the lack of a well controlled approach to study the critical properties at intermediate disorder strength for $d=1$, and at any disorder strength for $d>1$. Numerical methods, while very effective, have to face the challenge of faithfully extracting the bulk properties of systems whose behavior is often dominated by rare events. Numerical studies can correctly account for these rare events only by averaging the simulation results over a very large sample of the disorder statistics, or by using very large system sizes - both requirements involve a significant investment of computer time.

In this situation, a well-controlled experimental realization of dirty bosons with tunable chemical potential and/or interaction strength and/or disorder strength would allow a significant advancement of our understanding of this very complex quantum system. In the spirit of Feynman's analog quantum simulator idea  \cite{Feynman82}, a material (real or synthetic) exhibiting the dirty-boson quantum phase transition would provide an invaluable insight, not only in the static properties (which can be in principle reproduced by efficient numerical methods) but, most significantly, in the dynamic properties (which are instead challenging even for numerics). Intense investigations of dirty-boson physics have been or are being currently pursued in the context
of {\it e.g.} Helium absorbed in porous media  \cite{Crowelletal97}, granular superconductors  \cite{Sacepeetal11}, and ultracold atoms in disordered optical potentials  \cite{Fallanietal08, SanchezPalenciaL10}.

The focus of the preset review is, however, on magnetic insulators with gapped non-magnetic ground states. As quantum simulators for BEC and Bose Glass physics, these systems present numerous practical advantages. The  main source of magnetic interactions in insulators is superexchange, which is short-ranged and often correlated to structural features. This implies that real magnetic materials can be accurately described by simple spin models with just a few nearest-neighbor interaction constants. Helpfully, at low temperatures, lattice contributions are often negligible, and spins are the only degrees of freedom relevant to all thermodynamics and transport properties. In addition, an external magnetic field presents a very precise and well characterized experimental handle on the effective chemical potential of the bosonic quasiparticles. Last but not least, there are a number of particularly powerful experimental techniques, such neutron scattering, NMR and ESR, which allow a direct and quantitative measurement of the spin-spin correlation function. These measurements cover frequency (energy) scales of the typical exchange constant in real materials. Moreover, neutron scattering provides momentum transfers corresponding to typical lattice spacings, and NMR allows to inspect the average magnetic behavior at specific lattice sites.

\label{s.materials}
\subsection{Materials}

Most of the early work on quantum magnets was performed on transition metal oxides. These materials have features which have proved to be very important for seminal studies of quantum magnetism, but which can turn out as drawbacks in the present context. The energy scale of magnetic interactions, which is typically very large (on the order of hundreds of Kelvin), leads to magnetic transitions that might occur even at room temperature; but the flip side of such strong interactions is that these systems are then difficult to influence with magnetic fields, and any intrinsic magnetic
finite-temperature effect is hard to separate from those due to
lattice vibrations. As an example, for neutron spectroscopy high energies for magnetic interactions imply that
much of the spin excitation spectrum overlaps with phonon bands, and
is very difficult to separate from the latter. For our particular
purpose of modeling disorder physics, oxides have a disadvantage of
often being difficult to modify chemically without severely distorting
the crystal structure or influencing the electronic properties.
\begin{figure}
\begin{center}
\includegraphics[width=0.8\textwidth]{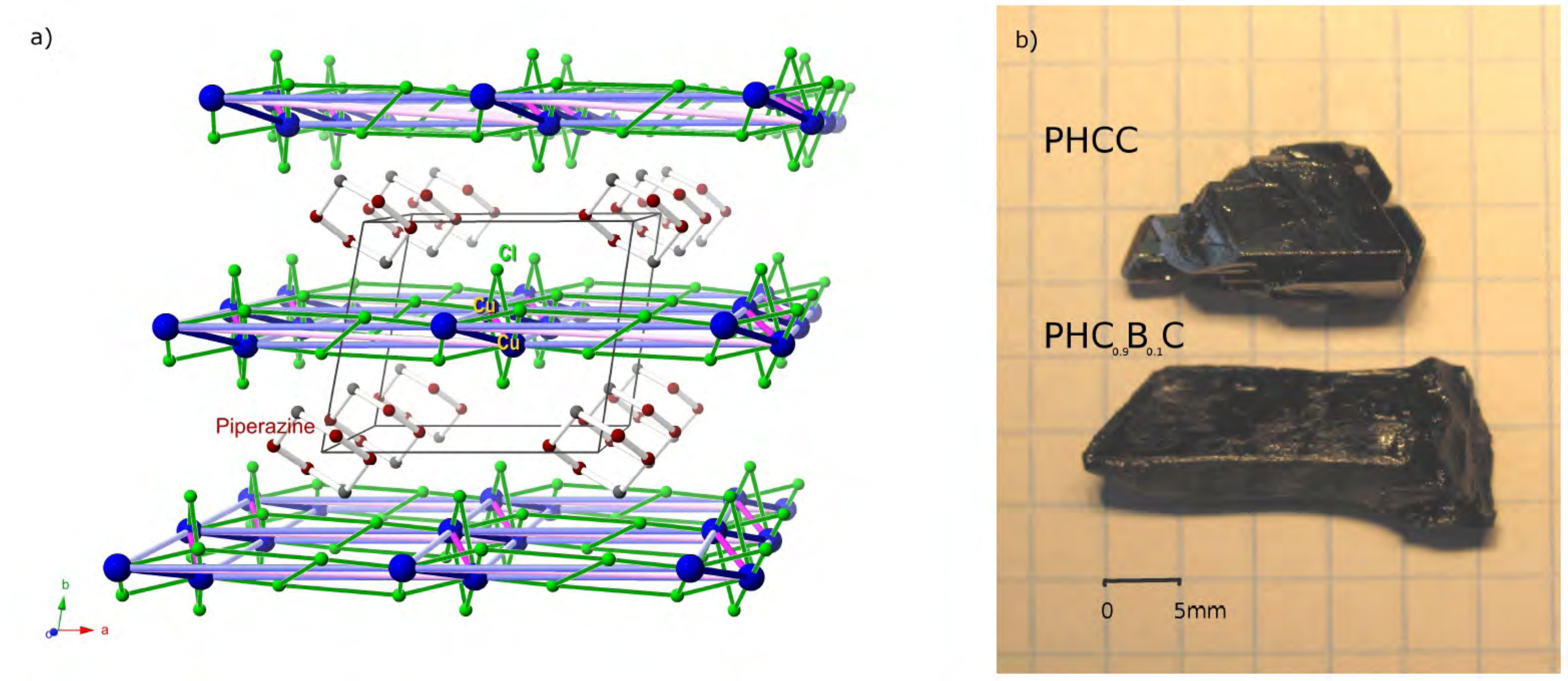}
\caption{(a) Crystal structure of one of the more complex organometallic quantum magnets, \PHCC. (b) large single crystals of \PHCC and \PHCX, $x=0.05$, grown from solution for calorimetric and neutron scattering studies  \cite{Yankova2012}.}
\label{f.PHCC}
\end{center}
\end{figure}

A true breakthrough came with the realization that excellent quantum spin systems are to be found among transition metal halogenides, particularly those that additionally incorporate organic ligands in the crystal structure. These compounds typically feature networks of spin-carrying transition metal cations connected via halogen bridges. The metal-halogen bonds are covalent in nature, and the resulting superexchange pathways can be unambiguously identified. Typical exchange constants are of the order of 1~meV (10~K).
Cu$^{2}$ and Ni$^{2+}$ are the most popular choices for $S=1/2$ and $S=1$ cations, respectively, and the (spinless) Cl$^-$ or Br$^-$ are the most frequently used anions. As an example, the structure of \PHCC, one of the more complicated materials of this type, is shown in Fig.~\ref{f.PHCC} (a).
In most cases, the organic ligands have a closed-shell electronic structure and are therefore not magnetic. They play the role of spacers in the crystal structure, and are to a good approximation uninvolved in the magnetism. Hundreds if not thousands of spin systems of this type have been identified. Among them are spin chains  \cite{Endoh1974,Hong2009}, spin ladders \cite{Watson2001,Masuda2006,Ruegg2008}, as well 2-dimensional \cite{Stone2001,Stone2006} and 3-dimensional spin networks \cite{Oosawa1999,Oosawa2001}, and some very interesting geometrically frustrated systems \cite{Garlea2008}.
Compared to oxides, these organic materials have numerous advantages.
The low energy
scale make them ideal for the study of field and finite-temperature
effects. It also enables the study of spin excitations using cold
neutron spectroscopy, which is a more precise and noise-free
technique compared to conventional thermal neutron scattering.
Finally, for many of the metallo-organic quantum magnets, it is quite
easy to grow large high-quality single crystals for experiments (Fig.~\ref{f.PHCC}b).
Unlike for oxides, sample preparation for halogenides is usually based on solution chemistry and therefore can be achieved with rather simple and
inexpensive equipment and relatively straightforward synthesis protocols  \cite{Yankova2012}.

\subsection{Magnetic BEC in transition metal halogenides}
It is among the transition metal halogenides that some of the most important experimental realizations of magnetic BEC were found. Here we shall mention only those materials that are currently being used as parent compounds for disordered systems realizing the Bose Glass  phase.

\begin{figure}
\begin{center}
\parbox[c]{0.4\textwidth}{
\includegraphics[height=0.35 \textwidth]{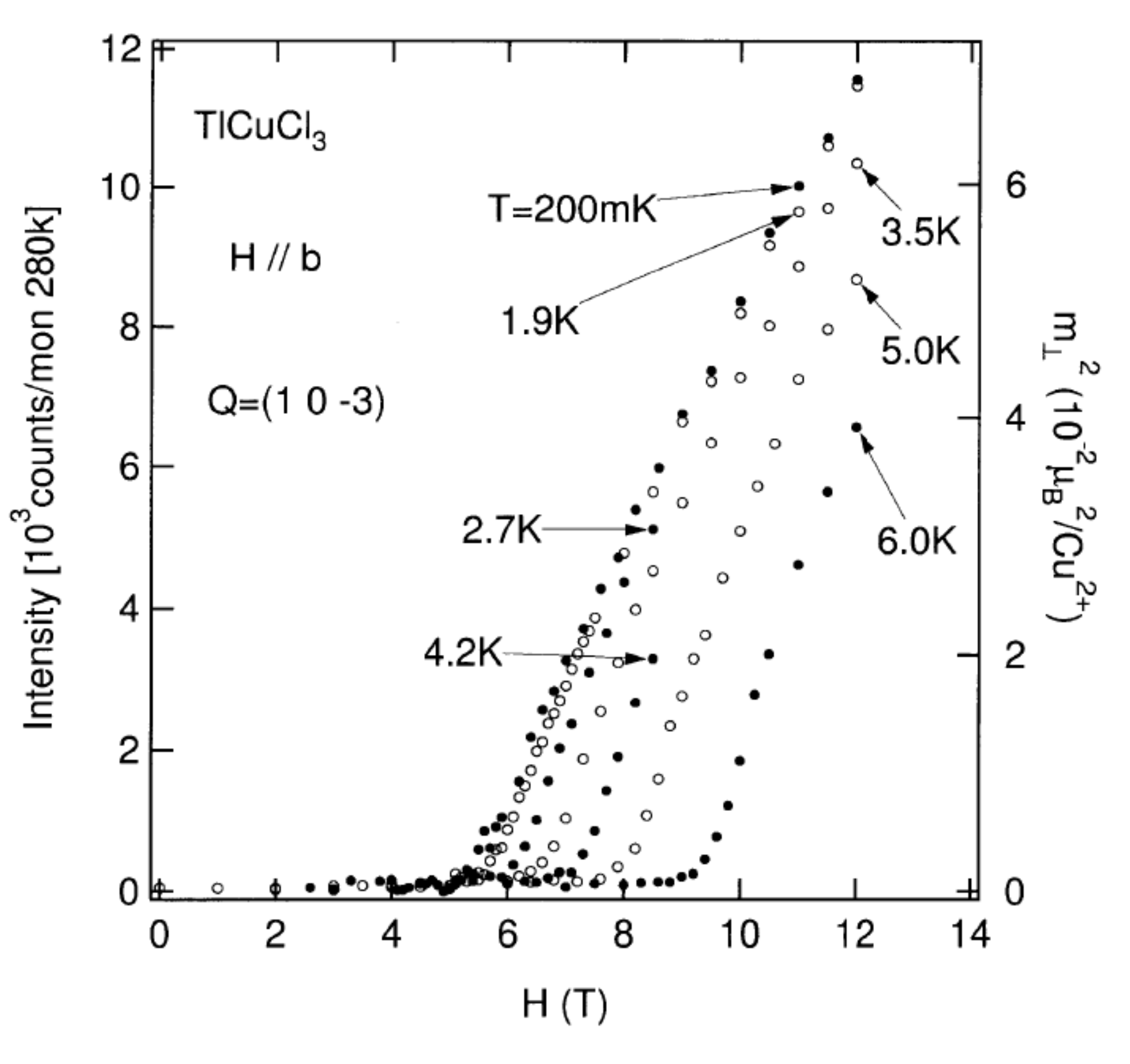}}
\parbox[c]{0.4\textwidth}{
\includegraphics[height=0.4 \textwidth]{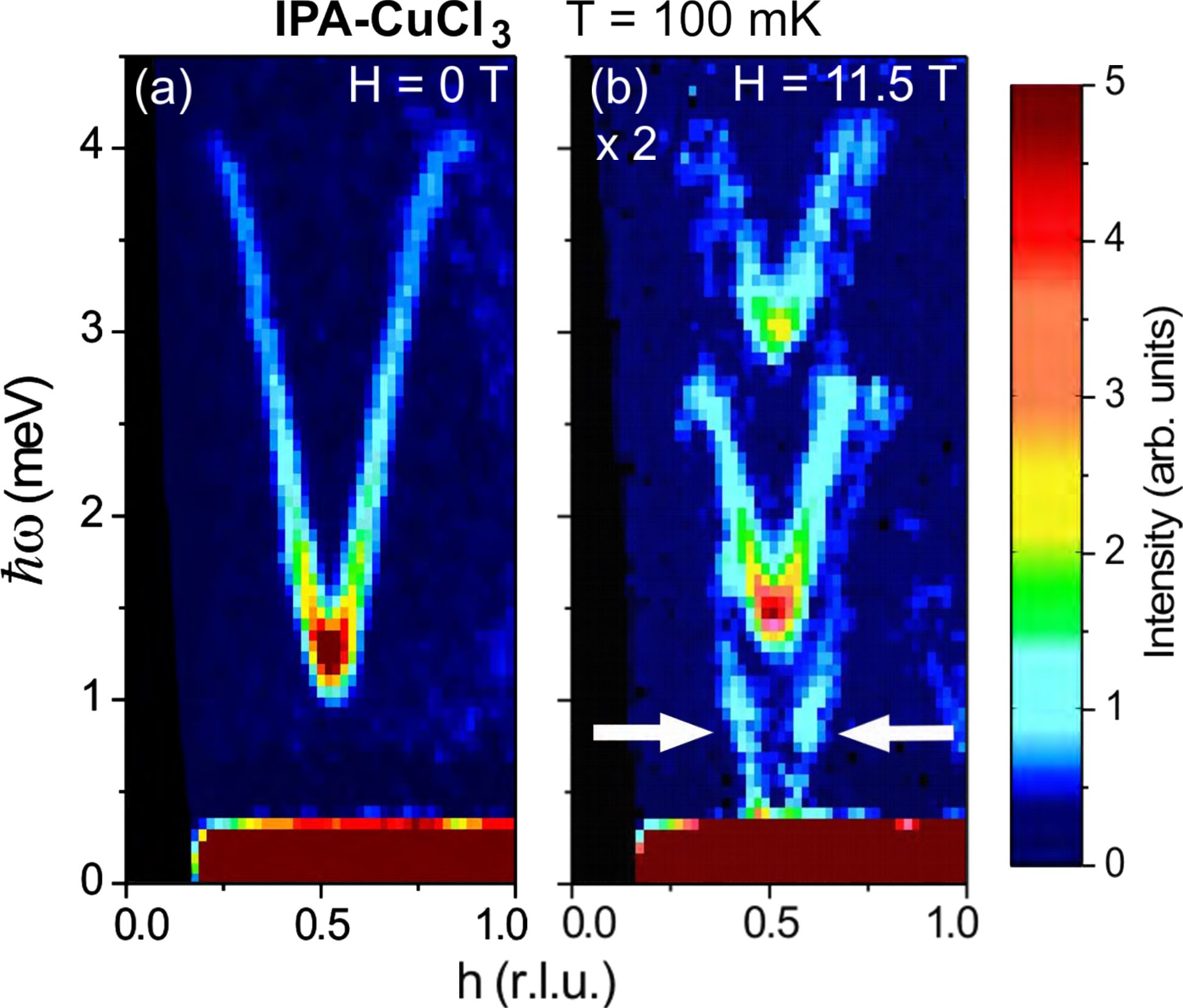}}
\caption{A key advantage of quantum magnets as simulators for BEC physics: neutron scattering can directly measure the BEC long-range order parameter (left, as for the case of \TCC, Ref.~ \cite{Tanaka2001}) and the Goldstone mode (white arrows on right, for IPACC  \cite{Garlea2007}).}
\label{f.TCC}
\end{center}
\end{figure}

Perhaps the most extensively studied is the $S=1/2$ dimer compound \TCC, where a field-induced magnetic BEC  was first reported in bulk measurements  \cite{Oosawa1999,Nikuni2000,Oosawa2001}, and later confirmed by neutron diffraction studies  \cite{Tanaka2001}. The critical field $H_c\sim 5$~T is easily accessible in experiments. The neutron scattering study was particularly important, as it highlighted one of the key advantages of magnetic insulators as quantum simulators for BEC: in these systems {\it the complex BEC order parameter is a directly observable quantity}. Indeed, the high-field BEC phase is characterized by spontaneous long-range order of the spin components transverse to the applied field. This transverse magnetization, written in complex form as $m_\bot=\langle S_x+ \ii S_y\rangle$,  corresponds to the complex wave function of the Bose condensate, while the intensity of magnetic Bragg peaks observed in neutron diffraction measurements is proportional to the square of its absolute value. Typical field dependencies of the thus measured BEC order parameter in \TCC  are shown in the left panel of Fig.~\ref{f.TCC}, following Tanaka {\it et al.}  \cite{Tanaka2001}.
Subsequent {\it inelastic} neutron scattering studies of \TCC probed the spin dynamics  \cite{Ruegg2002,Ruegg2003}, including the field dependence of the spin gap and the appearance of a gapless Goldstone mode. In magnetic BEC this mode is simply a spin wave in the magnetically long-range-ordered phase.

While \TCC is a pretty good BEC material, it has one subtle but important complication, namely a monoclinic crystal structure. Its low crystallographic symmetry ensures an intrinsic breaking of U(1) symmetry, that in the ideal case would only be spontaneously broken at the BEC transition. In fact, due to a residual anisotropy of exchange interactions  \cite{Glazkov2004}, the field-induced ordering transition is of the Ising universality class, as in numerous previously studied $S=1$ Haldane chain compounds \cite{Chen2001,Zheludev2004}. In particular, the high-field phase is actually gapped  \cite{Kolezhuk2004}. A related problem for TCC is the presence of Dzyaloshinskii-Moriya interactions. In fact, these interactions eliminate the phase transition altogether, replacing it with a crossover  \cite{Sirker2004}. Even though the anisotropy effects are rather small, they obviously affect critical behavior. In particular, the initially observed crossover exponent $\phi$ is about 0.47  \cite{Nikuni2000,Oosawa2001,Tanaka2001}, instead of the expected $\phi=2/3$. {Only in later measurements, performed in a deliberately chosen almost axially symmetric geometry, the BEC crossover exponent was recovered  \cite{Yamadaetal08}}. As will be discussed below, anisotropy takes on a special significance when one attempts to realize a Bose Glass state by introducing chemical disorder.

Two organic materials that are somewhat similar to \TCC, and that are relevant to our discussion of Bose Glass physics, are \IPA (IPACC, $H_c\sim 10$~T  \cite{Manaka1998}) and \PHCC (PHCC, Fig.~\ref{f.PHCC}, $H_c\sim 7.5$~T  \cite{Stone2001}). The former is a system of rather loosely coupled strong-rung spin ladders  \cite{Masuda2006}, while the latter realizes a quasi-2d spin network  with complex and somewhat geometrically frustrated interactions  \cite{Stone2001}. The field-induced BEC transitions have been extensively studied in both materials using thermodynamic measurements  \cite{Manaka1998,Stone2001,Stone2007}, neutron diffraction \cite{Garlea2007,Stone2007}, neutron spectroscopy  \cite{Zheludev2007,Stone2007}, and ESR  \cite{Manaka2001,Glazkov2012}.
For instance, the contrast between excitation spectra measured in the spin-gap and BEC phases of IPACC are shown in the right panel of Fig.~\ref{f.TCC}  \cite{Garlea2007}.  Here the arrows indicate the clearly visible Goldstone mode of the BEC transition.
The respective quasi-1d and quasi-2d nature of IPACC and PHCC leads to very peculiar features of the spin dynamics in zero field  \cite{Masuda2006,Stone2006-Nature,Zheludev2008,Nafradi2011} and across the BEC transition  \cite{Garlea2007,Zheludev2007}. Nevertheless the transition itself is a 3-dimensional one, and in many ways similar to that in \TCC. The two organic compounds are of triclinic symmetry and therefore are potentially subject to the same anisotropy problems  \cite{Nafradi2011,Glazkov2012} as \TCC.

\begin{figure}
\begin{center}
\parbox[c]{0.45\textwidth}{
\includegraphics[width=0.45 \textwidth]{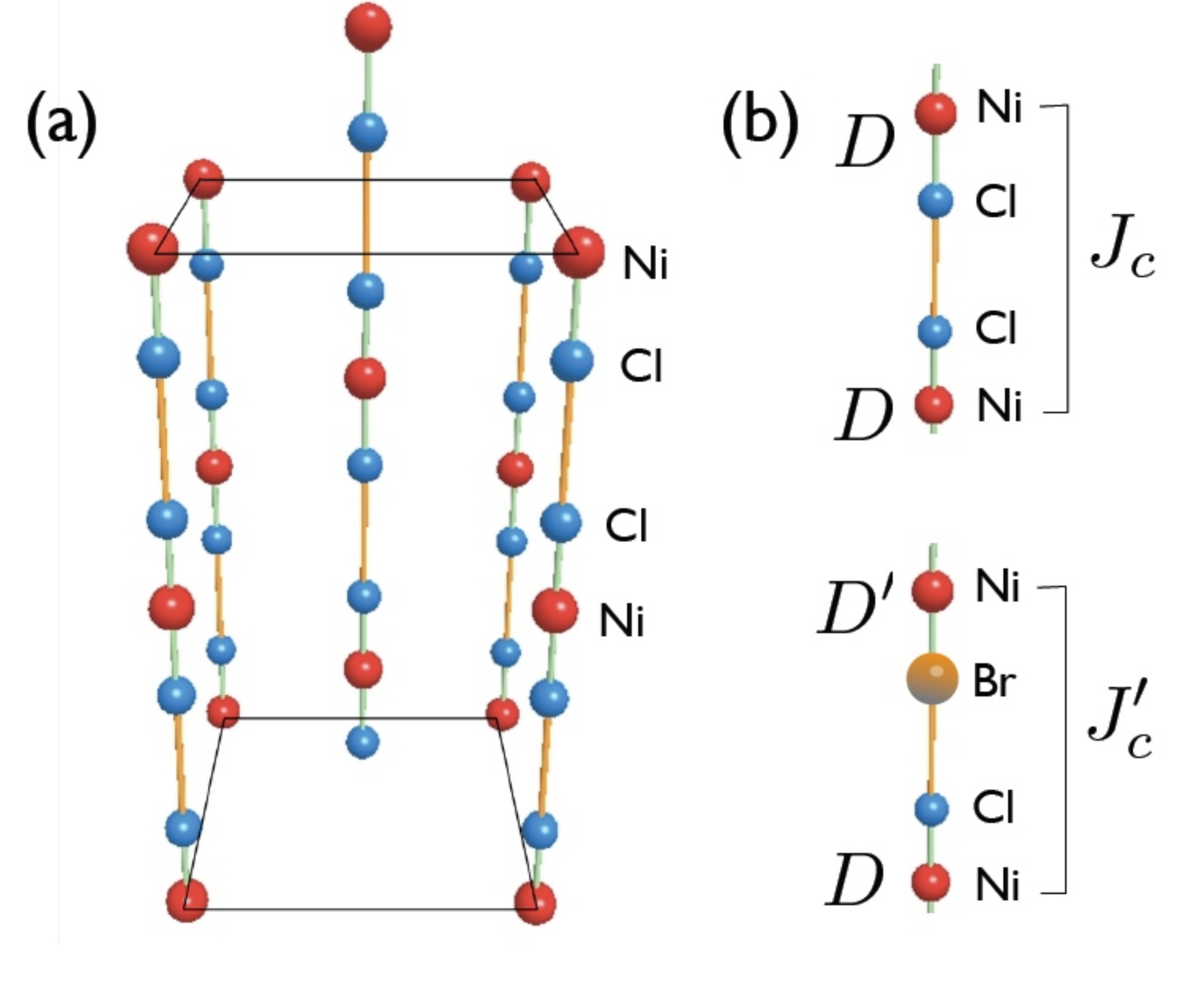}}
\parbox[c]{0.3\textwidth}{
\includegraphics[width=0.4 \textwidth]{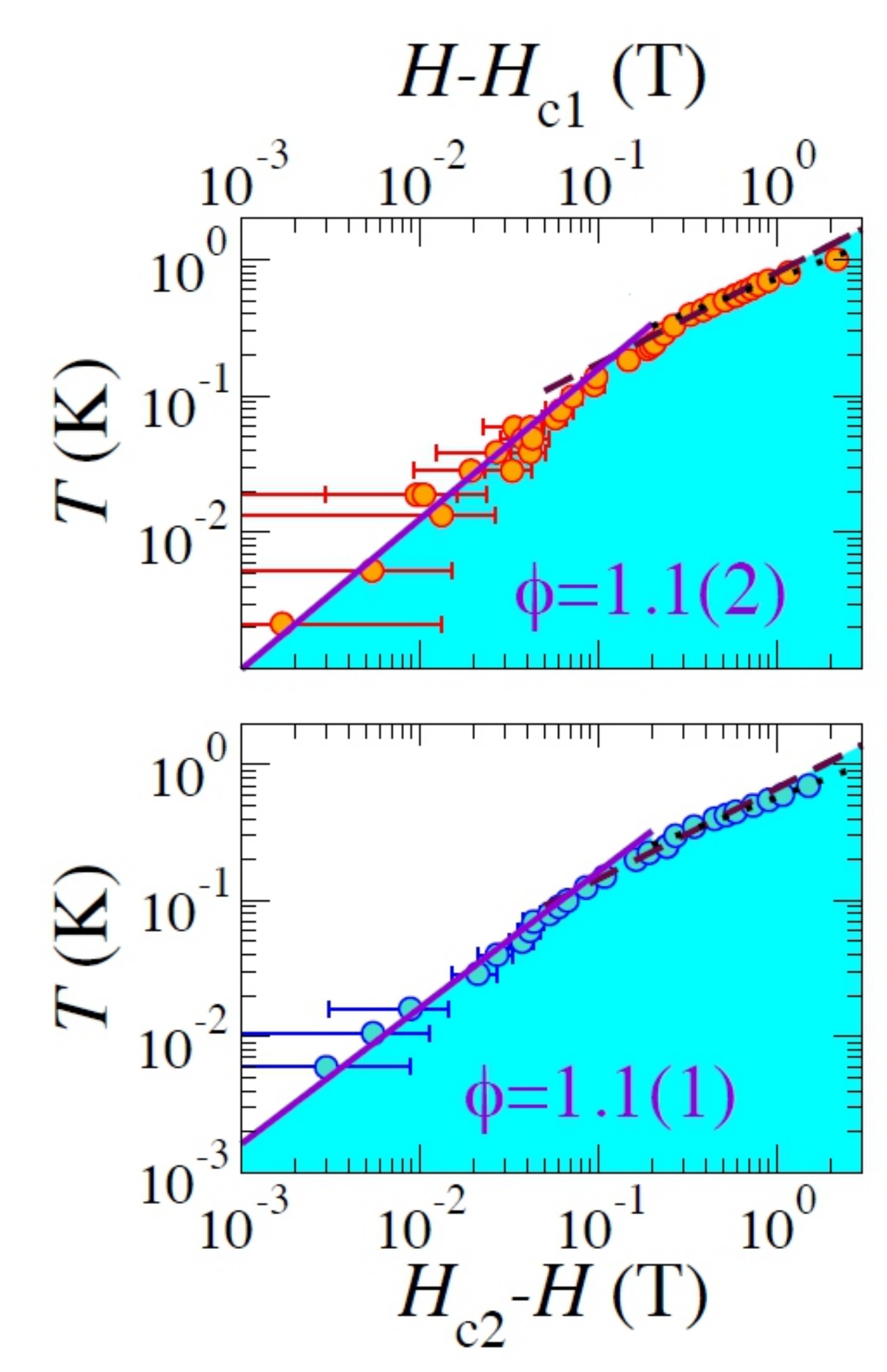}}
\caption{Left: (a) The body-centered tetragonal arrangement of $S=1$-carrying  Ni$^{2+}$ and the bridging Cl$^-$ ions in DTN. (b) effect of Br substitution on the coupling between Ni$^{2+}$ ions. Right: measured upper and lower phase boundary for the BEC phase in \DTNX, $x=0.05$ showing a low-temperature crossover to $\phi=1.1$. Both figures are adapted from Ref.~\cite{Yu2012}.}
\label{f.DTN}
\end{center}
\end{figure}

It is precisely because low crystal symmetry is potentially a problem for simulating a true BEC transition, that one material, namely \DTN (abbreviated DTN), stands out. It is a $S=1$ system, with a large easy-plane single-ion anisotropy. However, the crystal structure is tetragonal (Fig.~\ref{f.DTN}a), so the axial  symmetry is preserved as long as the magnetic field is applied along the unique $c$ axis. The ground state of each Ni$^{2+}$ ion is a non-magnetic singlet with $S_z=0$. A $c$-axis field drives the $S_z=1$ state below the singlet, which, due to exchange interactions between individual spins, results in long-range transverse ordering at $H_{c1}=2.1$~T  \cite{Paduan2004}. A second BEC transition from the fully saturate phase occurs at the upper critical field of $H_{c2}=12.6$~T, which is also readily accessible.  To date, DTN  is the cleanest known realization of magnetic BEC in a  spin-gap system  \cite{Zapf2006,Zvyagin2007}.

\subsection{Magnetic Bose Glass in transition metal halogenides}

\subsubsection{Chemical disorder}
The main reason that we are focusing on transition metal halogenide quantum magnets  is
that these materials turn out to be particularly advantageous for the study of disorder. Disorder is best introduced by a dilute and random
chemical substitution on the halogen site. A sketch of such a substitution is shown in Fig.~\ref{f.DTN}b. Unlike the much-studied
substitution on the magnetic sites, this type of chemical modification does not affect the spins themselves. What is modified, in a
spatially random manner, are the strengths of exchange interactions.
Superexchange via Br$^{-}$ is typically of the same sign, but several
times stronger than via Cl$^-$  \cite{Willett1986}, due to a different degree of covalency in the metal-halogen bond. By growing crystals with a certain  mixture of the two halogens in the starting reagents, one can thus obtain
materials that are very good realizations of random-bond spin
models. As discussed above, this is exactly what is needed to realize Bose Glass physics in dimer systems and many other spin-gap systems. Similar types of Hamiltonian disorder can be induced by  chemical substitution of other non-magnetic sites, such as Tl-K substitution in \TCX  \cite{Yamada2011}.

The larger radius of Br$^{-}$ ions relative to that of Cl$^{-}$, leads to local distortions around substitution sites and affects single-ion anisotropy. In anisotropic materials like DTN, this too is a source of random potential for the bosonic quasiparticles  \cite{Yu2012}. All these effects of chemical disorder (both random exchange and random anisotropy) can in some cases be very well quantified. This is exemplified by the impressive agreement between measurements on Br-substituted DTN, , \DTNX,  and first-principle quantum Monte Carlo calculations on a particular disorder model for this compound  \cite{Yu2012}, involving random bonds along the $c$-axis and correlated random single-ion anisotropies, but fully preserving the U(1) symmetry of the pure system.

\subsubsection{A compressible disordered phase}
\begin{figure}
\begin{center}
\parbox[c]{0.4\textwidth}{
    \includegraphics[height=0.4 \textwidth]{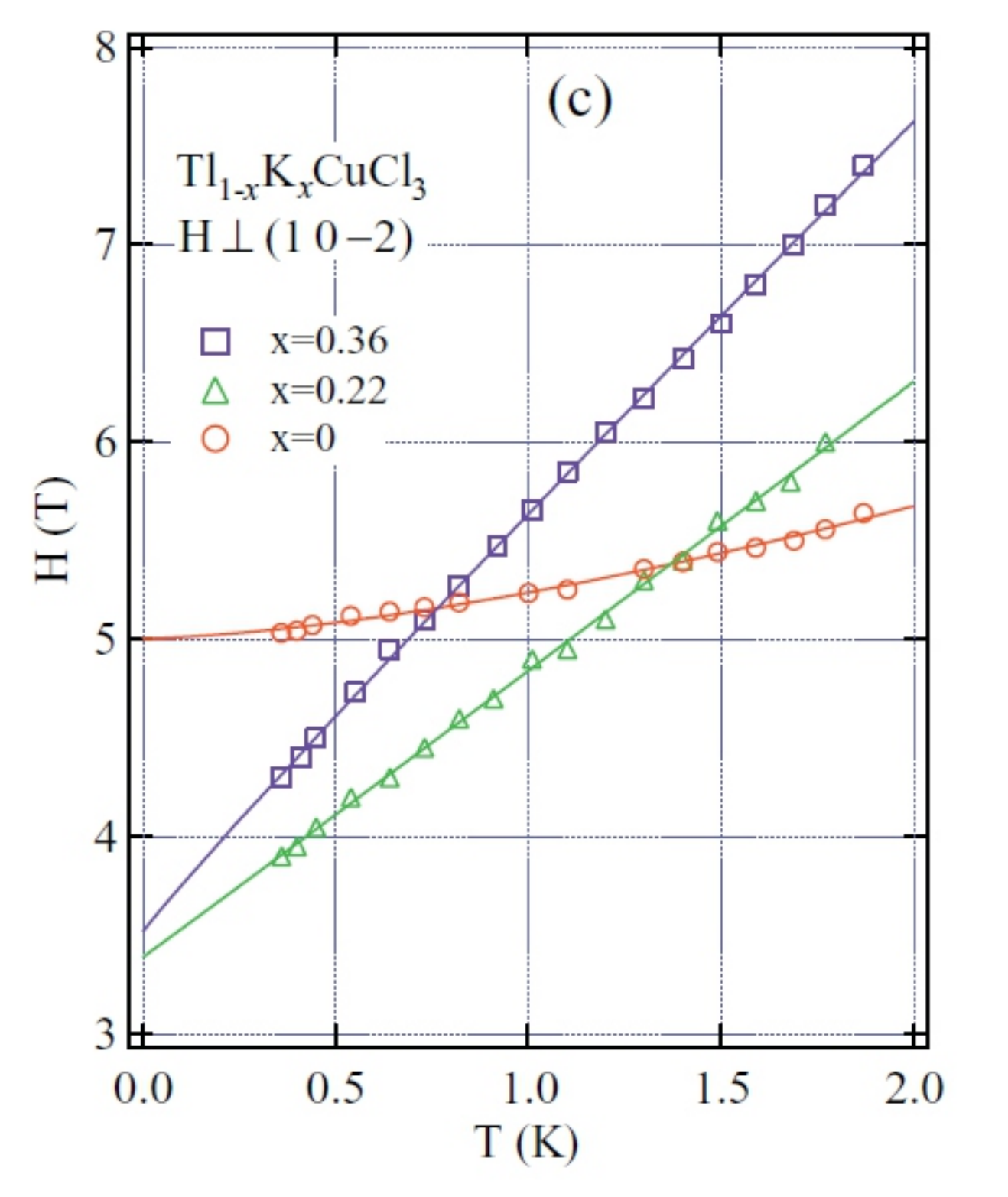}
}
\parbox[c]{0.5\textwidth}{
    \includegraphics[width=0.5\textwidth]{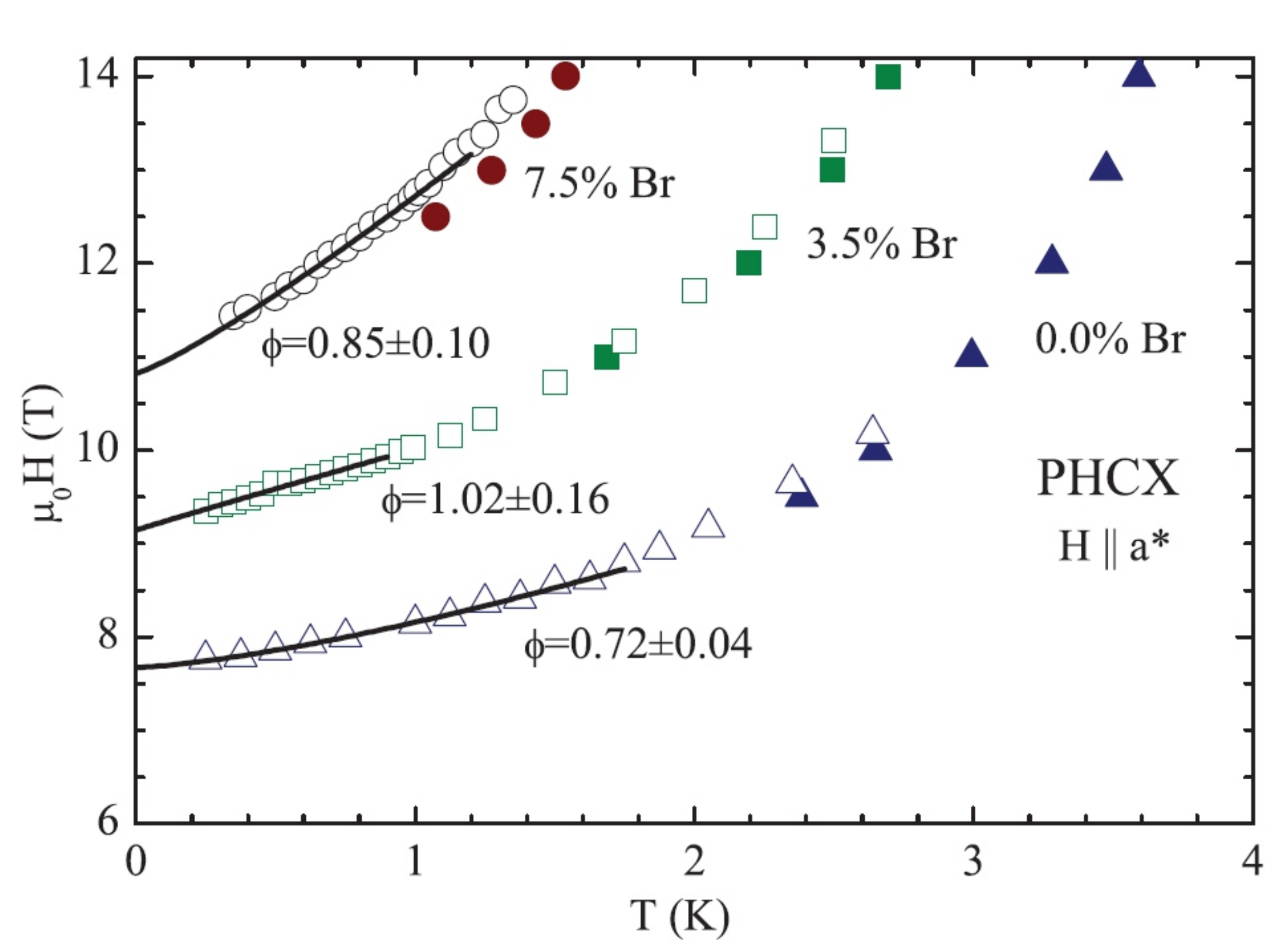}
}
\caption{Typical effect disorder on the BEC phase boundary in gapped quantum magnets. Left: K substitution on the Tl site in \TCX (Ref.~ \cite{Yamada2011}. Right: Br substitution on the Cl site in \PHCX  \cite{Huevonen2012}. }
\label{f.phaseboundary}
\end{center}
\end{figure}

\begin{figure}
\begin{center}
\parbox[c]{0.48\textwidth}{
\includegraphics[width=0.48 \textwidth]{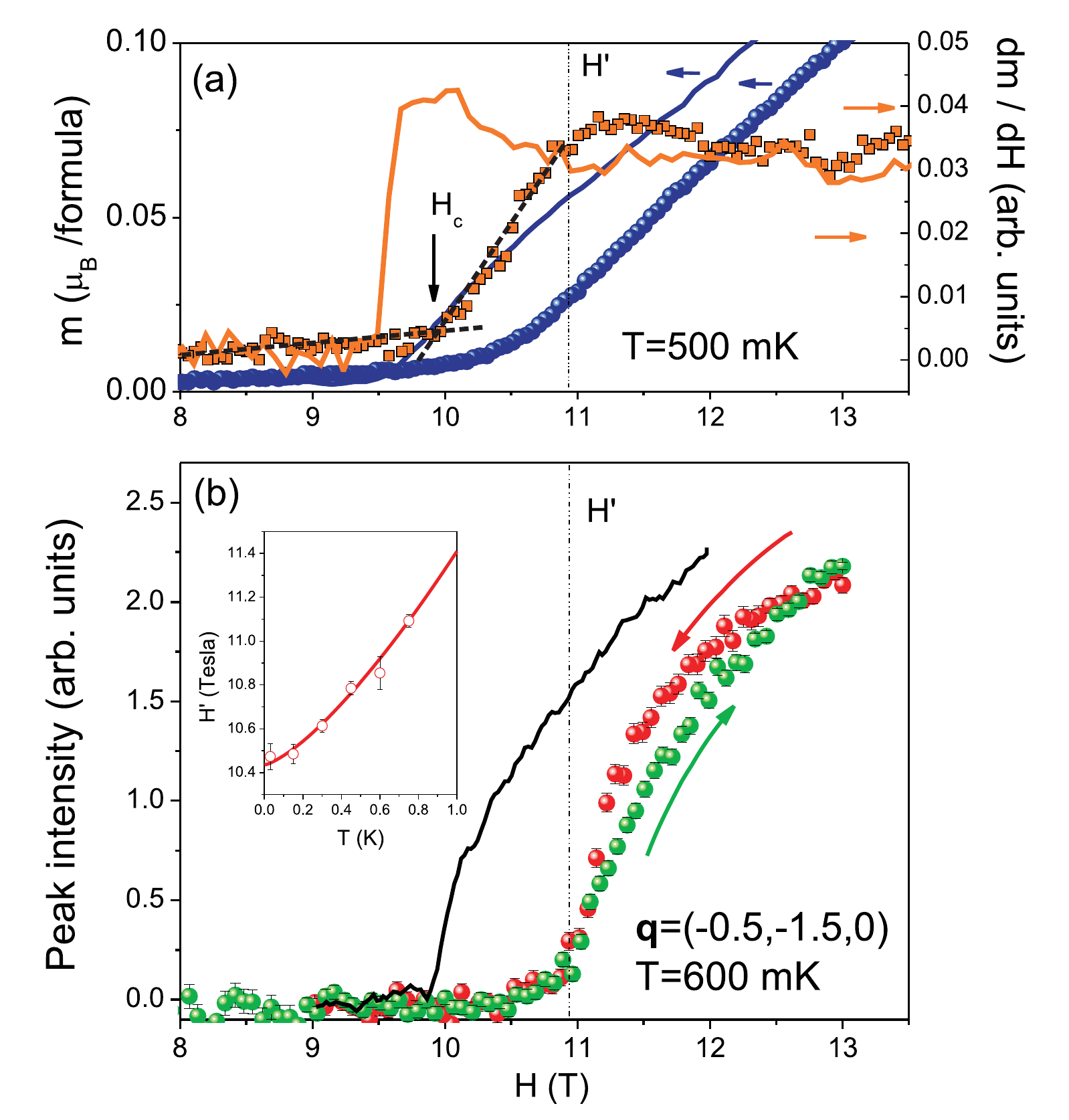}
}
\parbox[c]{0.48\textwidth}{
\includegraphics[width=0.48 \textwidth]{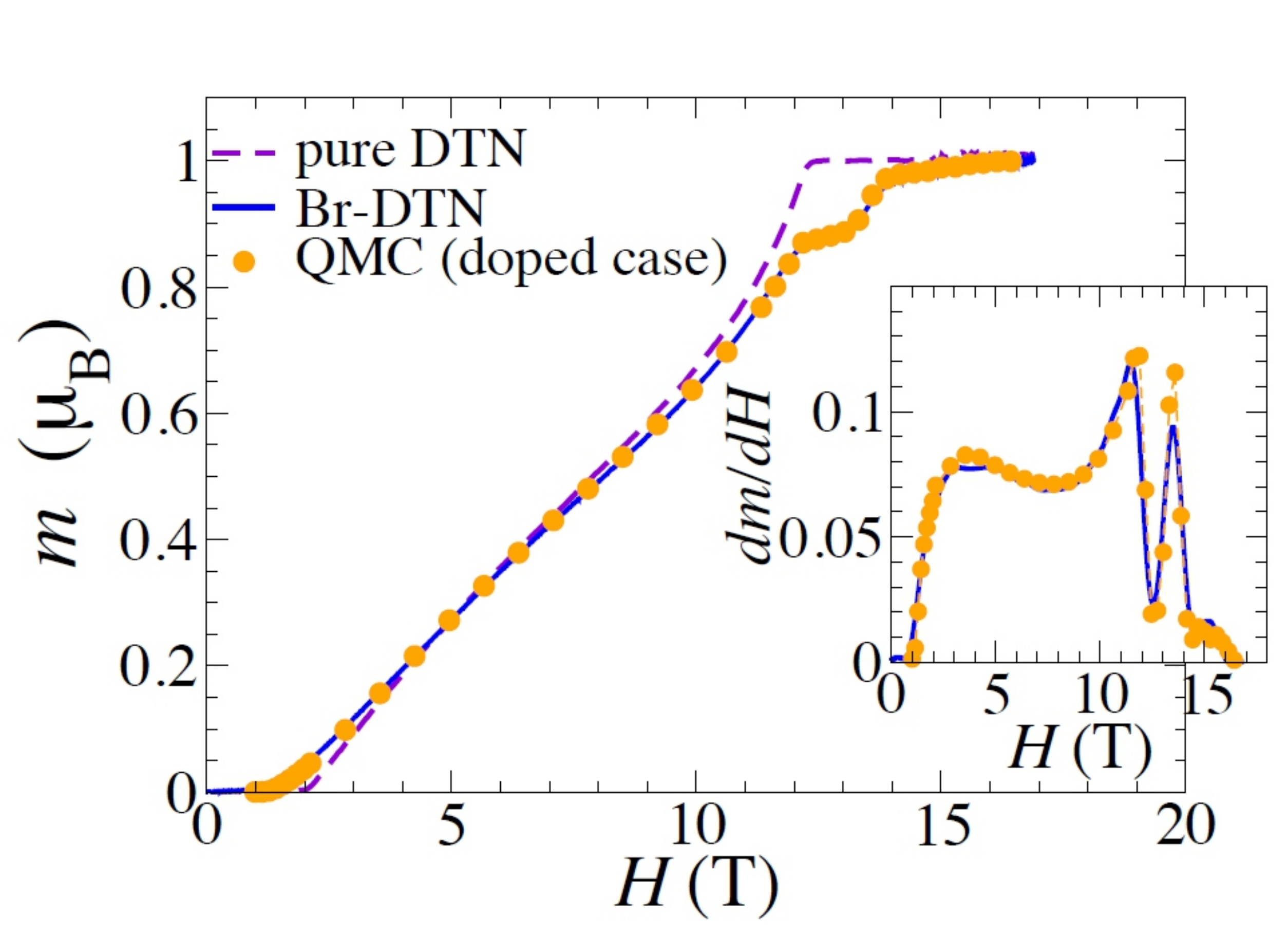}
}
\caption{Evidence for a compressible disordered phase (Bose Glass) in random gapped quantum magnets. Left: (a) Field dependence of magnetization (Boson density) and susceptibility (compressibility of the Bose gas) as measured in \IPA (solid lines) and \IPAX, $x=0.05$ (symbols). (b) Field dependencies of intensity of the antiferromagnetic Bragg peaks (square of the BEC order parameter) in the two respective materials. The data are from  \cite{Hong2010}. Right: Measured (solid lines) and calculated (symbols) field dependence of magnetization (boson density, main panel) and susceptibility (compressibility, inset) in Br-substituted DTN  (adapted from Ref.~\cite{Yu2012}).}
\label{f.mag}
\end{center}
\end{figure}

The most obvious effect of such a chemically induced modification of the Hamiltonian parameters is a shift of the BEC phase boundary on the H-T phase diagram. The effect can be quite dramatic, as shown in Fig.~\ref{f.phaseboundary} for the cases of PHCC and \TCX.
A more careful look reveals a more profound consequence of disorder. For magnetic BEC in disorder-free materials, the magnetic susceptibility $dM/dH$ remains essentially zero all the way to the critical field, and increases only in the transversely ordered  BEC phase. This corresponds to the gapped (Mott-insulator) phase being incompressible. For example, as illustrated in solid lines in the left panel of Fig.~\ref{f.mag}, for IPACC the susceptibility abruptly jumps at $H_c=9.8$~T (a), which coincides with the appearance of the magnetic Bragg peak  in neutron diffraction (b).
A key observation is that in all disordered gapped quantum magnets studied to date, non-zero longitudinal magnetic susceptibility already emerges in magnetic fields {\it below} the field of transverse long-range ordering. This {\it disordered compressible} state is a prime candidate for the Bose Glass phase. In the present example, for Br-substituted IPACC, \IPAX, $x=0.05$, the susceptibility starts to gradually increase  at roughly the same field as it jumps in the pure compound ( Fig.~\ref{f.mag}, left panel), but the antiferromagnetic Bragg peak only appears at a higher field, $H'=11$~T, where the susceptibility reaches a maximum. The field range $H_c<H<H'$ is interpreted as the domain of a magnetic Bose Glass. In contrast, in Br-substituted DTN, disorder shifts downwards the lower critical field to magnetic BEC, and the Bose glass phase extends from the critical field of long-range ordering all the way down to zero field  \cite{Yu2012}, as well as above the upper critical field, as discussed below.

In the vicinity of the upper critical field, the presence of a Bose Glass phase can be inferred from the magnetization curve. The latter may exhibit plateau-like features close to the transition, as is the case for Br-doped  DTN and IPACC. These plateaus are in fact pseudo-plateaus, retaining a finite albeit small slope, and they are due to a predominantly discrete (often bimodal) random distribution of  bond energies; the latter is indeed a good approximation for transition metal halogenides, where the  bond-strength-modulating effects of substitution ions are very local. The pseudo-plateau feature is then associated with the polarization of the spins linked {\it e.g.} by the weaker bonds, while the spins linked by the stronger bonds remain only partially polarized or unpolarized, hosting localized quasiparticles as discussed in Sec.~\ref{s.boseglass}.
Weak step features near the lower critical field were observed in \IPAX  \cite{Manaka2008,Manaka2009}, but the pseudo-plateaus are most prominent near the upper critical field of Br-substituted DTN, as shown in the right panel of Fig.~\ref{f.mag}. The remarkable agreement with QMC calculations once again emphasizes the validity of the theoretical model for disorder.

\subsubsection{The crossover exponent $\phi$.}
The observation of compressible disordered phases is  an important step, but the central physical issue remains that of the universality class of the BEC transition in disordered systems. The most easily accessible index in the experiments is the crossover exponent $\phi$ that can be measured using a variety of techniques used to trace the $H-T$ phase boundary. As mentioned in
Sec.~\ref{s.mBEC} and in the discussion above, for the BEC transition in 3-dimensional disorder-free systems we expect $\phi=2/3$, which increases to $\phi\sim 1.1$ in a Bose Glass to BEC transition in the presence of randomness. To date, the bulk of experimental data available supports this important theoretical prediction. Indeed, as can be see from Fig.~\ref{f.phaseboundary}, left panel, the phase boundary in \TCX  \cite{Yamada2011}, is essentially a straight line, despite some lingering controversy on this issue \cite{Zheludev2010comment,Yamada2011reply}. Similarly, a careful analysis of the phase boundary in $x=0.035$ \PHCX (Fig.~\ref{f.phaseboundary}, right panel) yields $\phi\sim 1.02(16)$  \cite{Huevonen2012}. The most accurate measurements to date are for Br-substituted DTN  \cite{Yu2012}. Measurements of magnetic susceptibility performed to the lowest temperatures reveal a crossover from $\phi\approx2/3$ at high temperatures to $\phi\approx1.1$  at temperatures  $\lesssim 0.1$~K, again in nice agreement with numerical simulations. The experimental results for both the lower and upper critical fields are shown in the right panel of Fig.~\ref{f.DTN}.

\subsection{The order parameter exponent and potential complications}
\begin{figure}
\begin{center}
\parbox[c]{0.4\textwidth}{
    \includegraphics[width=0.35 \textwidth]{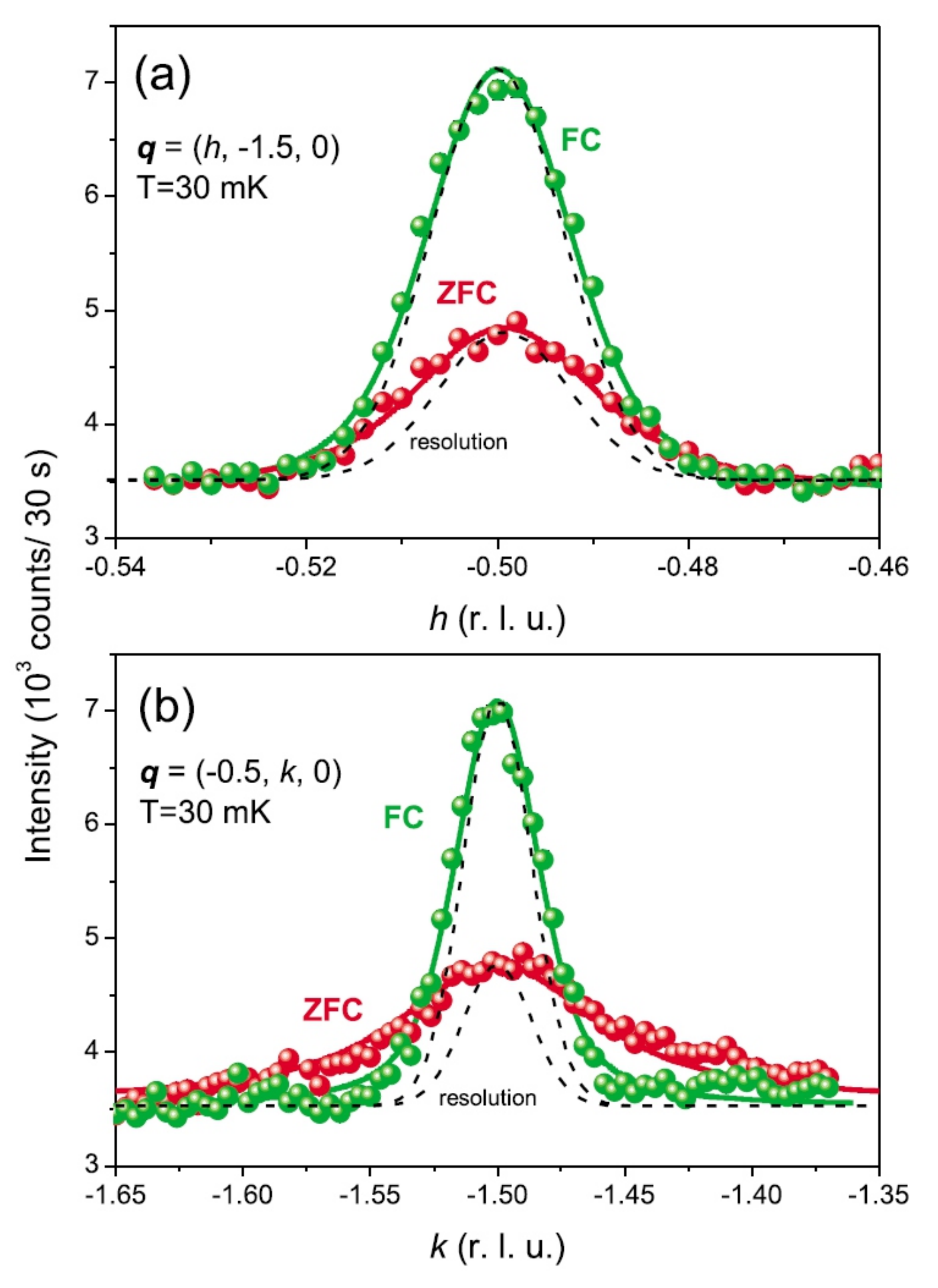}
}
\parbox[c]{0.5\textwidth}{
    \includegraphics[width=0.5\textwidth]{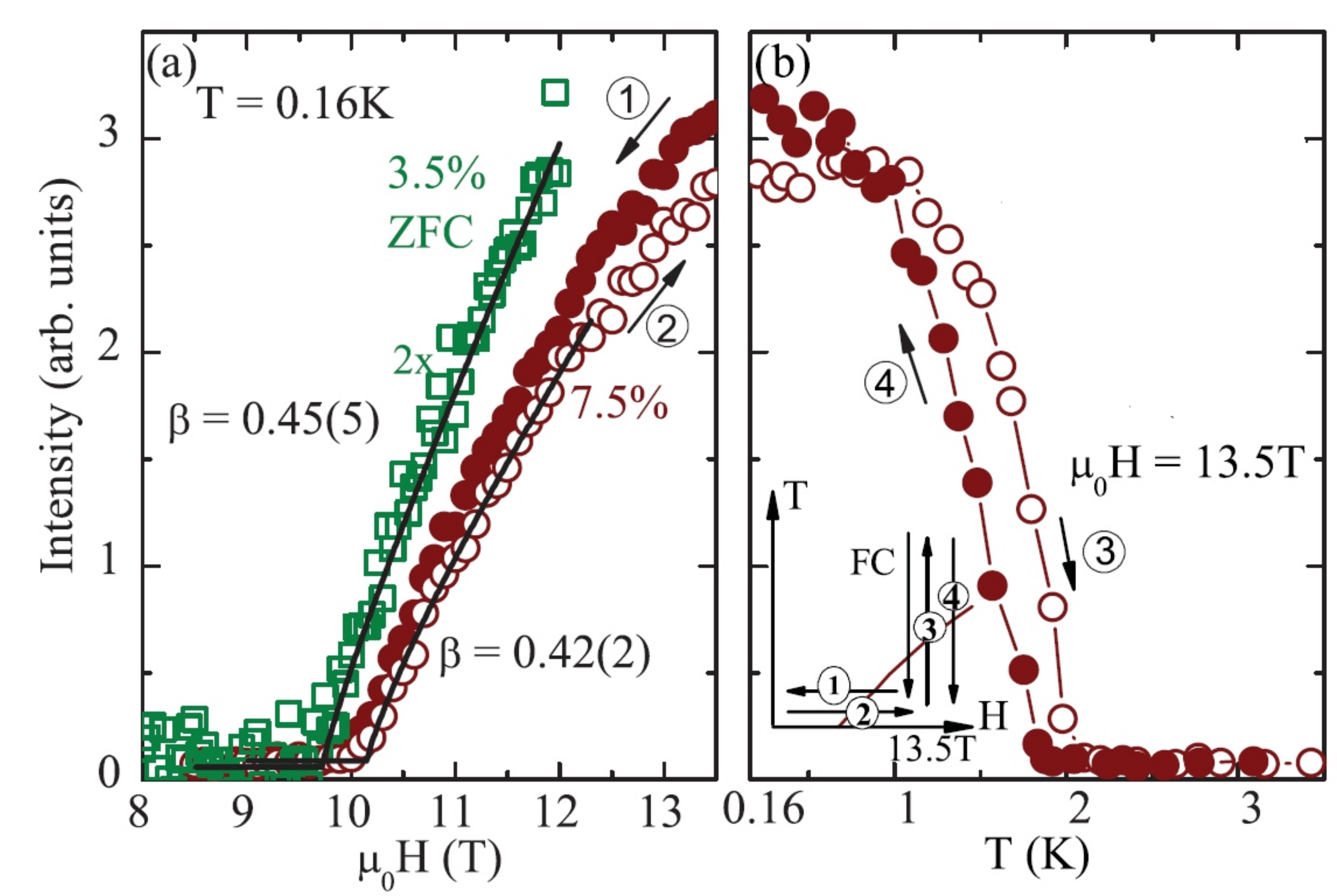}
}
\caption{Non-BEC behavior of elastic magnetic scattering in disordered triclinic quantum magnets. Left: Neutron scans across the magnetic Bragg peak  measured in Zero Field Cooled and Field Cooled \IPAX ($x=0.05$) at
T=30 mK in H=13 T applied field.
The magnetic peaks are broader than experimental resolution
(dashed lines),  \cite{Hong2010}. Right: Magnetic neutron diffraction peak intensity measured in \PHCX, (a) as a function of magnetic field and  (b) as a function of temperature. Open and
solid symbols correspond to zero field cooling and field cooling,
respectively. Squares (circles) correspond to $x = 0.035$ and  $x = 0.07$  \cite{Huevonen2012}.
}
\label{f.peaks}
\end{center}
\end{figure}

The most recent efforts were aimed at measuring the order parameter critical exponent $\beta$. As anticipated in Sec.~\ref{s.boseglass}, theory predicts that the mean-field value $\beta=0.5$ in disorder-free systems should be replaced by a much larger value $\beta \sim 0.95$ when disorder is introduced. Unfortunately, to date, there has been no clear verification of this prediction. In fact, the existing data for \IPAX  \cite{Hong2010} and \TCX  \cite{Yamada2011reply}, as well as much more careful measurements on \PHCX  \cite{Huevonen2012} suggest that the inclusion of disorder has little effect on the order parameter exponent. The same studies however, reveal some crucial features of the high-field phase in these materials that may be at the heart of the discrepancy. It was shown, for the cases of \IPAX \cite{Hong2010} and \PHCX \cite{Huevonen2012} that in these disordered compounds, unlike in their stoichiometric counterparts, the high field phase has only short-range order with a history-dependent correlation length.
For example, Fig.~\ref{f.peaks} (left panel) shows scans across the magnetic reflections in zero-field cooled and field cooled $x=0.05$ Br-doped IPACC samples, revealing a hugely different correlation lengths in the two cases. The right panel shows the field- and temperature- evolution of Bragg peak intensities in $x=0.035$ Br-doped PHCC, emphasizing their dependence on the sample's trajectory on the $H-T$ plane.

Clearly such behavior is inconsistent with a BEC state that should exhibit true long-range order and no history dependence. To understand it, we need to consider the more subtle and often undesired effects of chemical disorder. The first issue that cannot be ignored is that the random anisotropy introduced by chemical substitution need {\it not} be co-axial with the applied field, and thus should have a component in the plane of the spontaneous transverse magnetization. {\it Random anisotropy} of this type is known to have a drastic effect on thermodynamic phase transitions  \cite{Aharony1975}  and will undoubtedly have an effect on the quantum critical point. Even more disrupting is the fact that local strains around substitution sites will affect the orientation of the gyromagnetic tensor of the transition metal ion. In an externally applied magnetic field this will result in an effective {\it random field} acting on the spins. This random field will also have a transverse component, and is therefore a direct conjugate of the BEC order parameter. That random fields can totally disrupt long-range ordering in $d=3$, replacing it with a short-ranged ordered state is rigorously proven  \cite{Imry1975} and well documented for conventional magnets  \cite{Ferreira1985,Birgeneau1985}. In \IPAX and \PHCX  random field effects will be exacerbated by the fact that even the parent compounds may have some residual Ising anisotropy due to their triclinic structure,  given that a 3-dimensional Ising model in a random field does not show divergent Bragg peaks even in the ordered state.

Due to the tetragonal symmetry of the parent compound, in Br-doped DTN we expect these effects to be small. Indeed, it has been argued that axial symmetry is preserved when a Br$^-$ ion replaces a Cl$^-$ one, as shown in Fig.~\ref{f.DTN}b. Unfortunately, the symmetry is retained only {\it locally}. Except for the Ni$^{2+}$ ions that are strictly lined up along the affected chemical bond, strains propagating from each Br defect will distort the local ionic environments in a non axially symmetric fashion. Just due to this symmetry consideration, both random transverse anisotropy and random transverse field (in an applied external field) are to be expected. The key question, of course, is how strong and disruptive they may be. A hopeful sign, suggesting that they are, in fact, negligible, is that preliminary neutron diffraction experiments on $x=0.13$ \DTNX show no sign of short-range correlations or history dependence  \cite{Wulf2013unpublished}. The above-mentioned convincing agreement between experiment and theory  in other aspects of the problem for the $x=0.05$ material inspires further confidence. Careful measurements of the order parameter critical index and other exponents in DTN samples with different levels of Br doping are currently underway.

\section{Conclusions}
\label{s.conclusions}
In this paper we have discussed how disordered quantum magnets represent outstanding candidates for the quantum simulation of the physics of disordered bosons.
Fundamental theoretical questions in the field of disordered bosons are still open, such as the quantitative understanding of the onset of Bose condensation in a disordered environment. In particular a controlled treatment for the long-wavelength effective action, describing the transition from Bose glass to a superfluid condensate in dimensions higher than one, is still lacking; theoretical insight is mostly based upon a phenomenological scaling theory, which can lead to controversial results. In this context, controlled experimental realizations of dirty-boson physics, validated by extensive numerical simulations, can help resolve the controversies, and they can provide estimates for the critical and crossover exponents which a fundamental theory for the dirty-boson transition should be able to reproduce.

Understanding dirty-boson physics through its realizations in disordered gapped quantum magnets is very promising, but also quite challenging.  One of the main advantages of these systems is that BEC breaks the axial rotation U(1) symmetry  of the spins, and the resulting absence of quasi-particle number conservation is perfectly acceptable from a physical point of view. As a result, both the order parameter and relevant excitations are experimentally accessible. On the other hand the BEC phase is fundamentally related to the {\it spontaneous} breaking of the above cited symmetry, and hence it is vulnerable to all sorts of disruptive anisotropy effects, which might be enhanced by doping-induced disorder. This complication is by no means a show-stopper, but we are only now learning of its experimental implications. Fortunately, theoretical and numerical studies can provide invaluable and quantitative guidance in this endeavor. Future experimental and theoretical studies of magnetic quantum simulators for dirty-boson physics have the potential to reconstruct the quantum critical behavior at the dirty-boson transition via fundamental static observables such as the specific heat, order parameter and field-induced magnetization; and to investigate the evolution of the excitation spectrum across the transition, identifying the fundamental signatures of the phases connected by the transition at the level of the dynamical response functions.

\section{Acknowledgements}
A. Z. would like to acknowledge his present colleagues S. Gvasaliya, D. Huevonen, M. Haelg, W. Lorentz, G. Perren, D. Schmidiger, G. Simutis, M. Thede and E. Wulf for wonderful collaborations and exchanges of ideas. A. Z. also extends special thanks to his co-author on the present review, for his patience in explaining complex theoretical concepts to an experimentalist.
T. R. acknowledges continuing collaborations with R. Yu, S. Haas, V. Zapf, A. Paduan-Filho, L. Yin, N. Sullivan, F. Weickert, R. Movshovich, and all the authors of Refs.~\cite{Yu2012, Yuetal12-2} on the subjects of this review.
Support from a DOE INCITE award (T. R.) and the Swiss National Science Foundation (A. Z.), is gratefully acknowledged.
Both authors thank Thierry Giamarchi for collaboration and illuminating discussions.

\bibliographystyle{unsrt}
\bibliography{azbib,trbib}

\end{document}